%% file: models4.tex
\documentclass[letterpaper,11pt]{article}

\usepackage{epsfig}
\usepackage{epsf}
\usepackage{amsmath,amssymb}
\input{basic.tex}
\def\modelnum{twelve}
\newcommand{\NA}       {\mathsc{na}}

\begin{document}
\begin{titlepage}
\begin{center}
            \hfill    MCTP-03-55 \\
            \hfill    UPR-1059 \\
            \hfill    MADPH-03-1364 \\
            \hfill    hep-ph/0312248
\end{center}
\vspace{0.2cm}

\begin{center}
{\Large \bf Relating Incomplete Data and Incomplete Theory}
\end{center}
\vspace{0.2cm}

\begin{center}
P.~Bin\'etruy$^{1}$, G.~L.~Kane$^{2}$, Brent~D.~Nelson$^{3}$,
Lian-Tao~Wang$^{4}$ and Ting~T.~Wang$^{2}$
\end{center}

\begin{center}
$^{1}${\it Laboratoire de Physique Th\'eorique,\\

Universit\'e Paris-Sud, F-91405 Orsay, France \\

and APC, Universit\'e Paris 7, College de France, F-75231 Paris,
France}
\end{center}

\begin{center}
$^{2}${\it Michigan Center for Theoretical Physics, Randall
Lab.,\\

University of Michigan, Ann Arbor, MI 48109}
\end{center}

\begin{center}
$^{3}${\it Department of Physics, David Rittenhouse Lab.,\\

University of Pennsylvania, Philadelphia, PA 19104}
\end{center}

\begin{center}
$^{4}${\it Department of Physics,\\

University of Wisconsin, Madison, WI 53706}
\end{center}

\vspace{0.4cm}

\begin{quotation}\begin{small}
Assuming string theorists will not soon provide a compelling case
for the primary theory underlying particle physics, the field will
proceed as it has historically: with data stimulating and testing
ideas.  Ideally the soft supersymmetry breaking Lagrangian will be
measured and its patterns will point to the underlying theory. But
there are two new problems.  First a matter of principle: the
theory may be simplest at distance scales and in numbers of
dimensions where direct experiments are not possible.  Second a
practical problem: in the foreseeable future (with mainly hadron
collider data) too few observables can be measured to lead to
direct connections between experiment and theory. In this paper we
discuss and study these issues and consider ways to circumvent the
problems, studying models to test methods. We propose a
semi-quantitative method for focusing and sharpening thinking when
trying to relate incomplete data to incomplete theory, as will
probably be necessary. \end{small}\end{quotation}

\begin{footnotesize}
$^{*}$This work was supported in part by grant number
DE-FG02-95ER40893.\end{footnotesize}

\end{titlepage}

\pagebreak
\section{Introduction}

We are living in an idea-rich era, when it comes to supersymmetric
(SUSY) model-building. While the generic ``problems'' of SUSY
models are often emphasized -- the problem of dynamical SUSY
breaking, the flavor problem, the problem of CP-violating phases,
the $\mu$ problem, etc. -- these problems are typically exhibited
only for the sake of motivating a new solution. In fact most, if
not all, of the generic problems of SUSY models have been
``solved'' several ways. Yet no ``Supersymmetric Standard Model''
exists. We believe the reason for the absence of such a model is
not that none of the solutions mentioned above are satisfactory.
Instead it is that most models of beyond the Standard Model
physics based on supersymmetry only address one or two of these
problems, with the remaining unaddressed problems leaving behind
implicit large hierarchies and fine-tunings. As such these models
cannot be considered realistic in the sense that they (of
necessity) fail to explain a large segment of the world we observe
experimentally.

This is a serious weakness given that even in the absence of
direct evidence for superpartners we actually know a great deal
about any possible supersymmetric model. The measured parameters
of the Standard Model alone provide severe constraints on any
putative supersymmetric Standard Model (SSM). Collider data,
low-energy experiments and cosmological observations all provide
further constraints. Yet the unrealistic nature of most SUSY
models in the literature implies that only a small subset of this
data is ever brought to bear on a given model. This deficiency
will become even more obvious in the data-rich era of
supersymmetry which we believe is at hand. One might contend that
as soon as superpartners are directly observed models will rapidly
improve. But consider (for example) a trilepton signal occurring
at the LHC. Such an observation would be very exciting. It would
give us some, but limited, information about chargino and
neutralino masses -- and even tell us something about the nature
of dark matter. But it would not allow the measurement of SUSY
Lagrangian parameters or $\tan\beta$. What it would tell us is for
the most part already assumed in the model-building that has
occurred thus far.

A supersymmetric Standard Model is likely to emerge only when we
are able to take full advantage of the current and future data and
understand how this data might connect to a fundamental theory
such as string theory. This, in turn, is likely to require a
change in strategy from all elements of our community. The
standard approach of phenomenologists has been to assume that
reconstructing a fundamental theory involves first measuring --
often to a high degree of precision -- the parameters of the soft
supersymmetry-breaking Lagrangian which can then be connected to a
supersymmetric Standard Model at a subsequent stage~\cite{review}.
The standard approach of SUSY model-builders has been to assume
that each of the issues that a SUSY standard model must address
can be treated in isolation, with ingredients brought together and
incorporated into a single model with ease at some later time. The
standard approach of string theorists has been that any use of
string constructions to guide this model building is premature,
given our lack of a complete understanding of the space of all
possible string theories. We believe that none of these
assumptions is well justified and that enough information is
available now, or may soon exist, to allow progress.

Perhaps past experience can provide some support for this view.
When the Standard Model was formulated only a little was known and
some only tentatively: the hadron spectrum, the existence of
quarks and two neutrinos, currents were vector and axial vector,
fermions were chiral, weak interactions were weak and early
scaling in deep inelastic scattering. Theoretically the framework
of gauge theories and the renormalizability of weak interactions
were in place. Similar kinds of experimental information and
theoretical structures are in place today. Some existing models
are rather comprehensive and describe a lot of phenomena, yet they
are not elevated to the position of supersymmetric standard models
by the community. That may be because they involve fine-tunings
and/or do not have a clear connection to an underlying theory such
as string theory.

Thus we propose an improved way of thinking about how high energy
models and experimental observations are to be connected in
anticipation of the data-rich era expected to come. This approach
is based on several principles. First, SUSY models must be made
more realistic -- efforts must be made to move beyond toy models
to more holistic ones. This means models should begin to address
all of the issues that a supersymmetric Standard Model might be
expected to explain. Second, creative thinking is needed in
identifying what we have called ``inclusive signatures.'' This
means finding collider and non-collider observables which are both
actual observables in existing and forthcoming experiments and
which more directly probe the key features of supersymmetric
Lagrangians. Third, we recommend finding ways in which observables
from all arenas -- low-energy, collider, cosmology, etc. -- can be
used in conjunction to get to the most likely paradigms as quickly
as possible, and we propose a new method for doing so. This
approach supplements and complements the more systematic bottom-up
one of going from data to the Lagrangian at low energies to the
high scale Lagrangian to the underlying (and presumably string)
theory by helping to proceed with incomplete data.

In this paper we look at the state of the current approach by
studying inclusive signatures and theoretical features
of~\modelnum~benchmark points from~\modelnum~different models
drawn from the current literature. The precise nature of these
models is largely irrelevant to the new way of thinking we are
proposing here, though we provide a very brief description of each
model in Section~\ref{sec:models}; many readers can skip this
section. We then use this survey as a backdrop in
Section~\ref{sec:obs} for a discussion of how current and future
data can provide considerable discriminatory power even before
precision measurements of soft Lagrangian parameters are made. In
Section~\ref{sec:pheno} we introduce new ways of confronting the
different theories with experimental data that seeks to maximize
the power of incomplete data. Experimentalists may wish to read
Sections~\ref{sec:disc},~\ref{sec:IPO} and~\ref{sec:exp} first. We
then make suggestions based on this new approach for model
building and formal theory in Sections~\ref{sec:mod}
and~\ref{sec:string}. In a concluding section we speculate on how
a supersymmetric Standard Model will emerge and comment on how our
suggestions might help bring that about.

\section{A Sampling of Models}
\label{sec:models}

Our goal is to understand how the three elements we described
above -- formal theory, model building and phenomenology -- work
together to provide interpretations of observations and how this
partnership can be made more efficient. The product of the
combined efforts of these arenas, with input from experiments,
should eventually be a supersymmetric Standard Model. Our own
theoretical prejudice orients us towards models which are more
likely to have a string-theoretic origin. However we here
present~\modelnum~different models which span a wide spectrum of
ideas. No effort was made to be absolutely comprehensive -- nor do
we make judgements here about the relative merits of any one
model. Our one requirement is that the models admit some limit in
which they can be treated as some form of MSSM model. All of the
models we consider are designed to be consistent with the apparent
unification of gauge couplings at some high scale. Thus we will
not consider, for example, TeV-scale string models, models of
(large) universal extra dimensions or little Higgs models.

The SUSY models described below primarily concern themselves with
the parameters of the soft supersymmetry-breaking Lagrangian and
hence the pattern of superpartner masses.  We reiterate that the
actual models chosen, as well as the specific point in the
parameter space that we study, are largely irrelevant for the rest
of the argument. The purpose of these summaries is to show the
diversity of ideas current in the literature. Below we will
briefly describe each model's features before discussing the
observational consequences and how one might distinguish between
them in Section~\ref{sec:obs}.

\vspace{0.1in}
\noindent \textbf{Model A: Generic mSUGRA.}

The minimal supergravity (mSUGRA) model is defined by a universal
soft supersymmetry-breaking gaugino mass $m_{1/2}$, a universal
scalar mass $m_0$, a universal trilinear coupling $A_0$, as well
as the electroweak parameters $\tan\beta$ and ${\rm sgn}(\mu)$. It
is the simplest and most often studied model of supersymmetry
breaking soft terms. We will pick a point in the low mass region
of the allowed parameter space as a baseline for comparison with
the other models to follow. The point we choose is Point~B of
Battaglia et al.~\cite{Battaglia}, slightly adjusted to achieve a
reasonable Higgs mass as in the first of the Snowmass Points and
Slopes~\cite{Snowmass}: point SPS~1a. This point is given by $m_0
= 200 \GeV$, $m_{1/2} = 250 \GeV$, $A_0 = -800 \GeV$,
$\tan\beta=10$ and positive $\mu$.

\vspace{0.1in}
\noindent \textbf{Model B: Hyperbolic/focus point mSUGRA.}

There exists a locus of points in the mSUGRA parameter space for
which large radiative corrections to the one-loop effective
potential result in a small value of $\mu$ once the electroweak
symmetry breaking (EWSB) constraint is imposed. For such
``hyperbolic branch'' points, the cancellations necessary to
achieve a Z-boson mass of $M_Z = 91 \GeV$ are greatly
reduced~\cite{ChChNa98}. Viewed in this way these points, which
tend to involve large values of the (universal) scalar mass, may
be considered ``natural'' in some sense. This view is strengthened
by the observation that the running of the scalar soft mass
$m_{H_u}^{2}$ exhibits a focus point behavior at low
energies~\cite{FeMo00,FeMaMo00a,FeMaMo00b}. Thus the large scalar
mass region of mSUGRA where the $\mu$ parameter is rapidly driven
towards zero has come to be known as the ``focus point'' region --
though the focus point behavior is operative throughout the mSUGRA
parameter space.

We choose to define this region as the set of points for which
$\mu \leq 250 \GeV$ with $m_0 \geq 1 \TeV$. The precise location
of this space for a given value of $m_{1/2}$ is theoretically
uncertain and varies from one study to another depending on the
analysis techniques employed. For our study we find an example
with $\mu = 210 \GeV$ for $m_0 = 2150 \GeV$ and $M_{1/2} = 300
\GeV$ with vanishing A-term for $\tan\beta = 10$. This point is
similar to point SPS~2~\cite{Snowmass} from the Snowmass Points
and Slopes.

\vspace{0.1in}
\noindent \textbf{Model C: Minimal Gauge Mediation.}

The gauge mediation model is characterized by a messenger sector
which transmits the information of supersymmetry breaking from a
hidden sector to the observable sector through gauge
interactions~\cite{GiRa99}. In its minimal form this sector is
assumed to comprise of N families of fields in a vector-like
$\mathbf{5} + \mathbf{\oline{5}}$ representation of $SU(5)$ to
preserve gauge coupling unification. The messengers are assumed to
have universal (supersymmetric) mass $M_{\rm mess}$ and a mass
splitting between scalars and fermions determined by the parameter
$F$ such that observable sector soft masses are determined by the
ratio $\Lambda = F/M_{\rm mess}$. These masses are presumed to be
determined at the scale $\Lambda_{\UV} = M_{\rm mess}$. We will
take as a representative point in the minimal gauge-mediated
parameter space the case of the Snowmass point
SPS~8~\cite{Snowmass} with $\Lambda=100 \TeV$, $M_{\rm mess} =
\Lambda_{\UV} = 200 \TeV$, $N=1$ and $\tan\beta=15$.

\vspace{0.2in}
\noindent \textbf{Model D: Minimal Anomaly Mediation.}

This is the original anomaly mediation model based on K\"ahler
potentials of the sequestered sector form as suggested
in~\cite{GiLuMuRa98,RaSu99,PoRa99}. The problem of tachyonic
slepton masses arising first at two loop order is addressed
through the addition of a universal contribution to scalar masses
of undetermined origin. The phenomenology of this model was
investigated in~\cite{GhGiWe99} and later incorporated into the
Snowmass Points and Slopes as the point SPS~9~\cite{Snowmass}.
This is the point in the parameter space that we will investigate.

\vspace{0.1in}
\noindent \textbf{Model E: Anomaly Mediation with Ancillary U(1).}

In this model the slepton problem of anomaly mediation is overcome
by including the effects of an additional $U(1)$ D-term on scalar
masses. This ``ancillary'' $U(1)$ is assumed to be anomaly-free
using only the particle content of the MSSM. This restricts the
possible $U(1)$ charge assignments which can be parameterized by
two rational numbers~\cite{MuWe03}. Requiring that slepton
mass-squareds be positive further restricts these choices. We will
investigate the properties of an example in which the D-term
contribution to scalar masses has a magnitude that is roughly
three times the size of the typical scalar mass ($\eta = 3$ in the
language of~\cite{MuWe03}). We choose charge assignments such that
$Q_E = 1$ and $Q_Q = -1/10$ with a gravitino mass of $m_{3/2} = 79
\TeV$.

\vspace{0.1in}
\noindent \textbf{Model F: Heterotic Strings with K\"ahler
Stabilization.}

The next two models involve weakly coupled heterotic string theory
and are derived from the two most commonly employed methods for
stabilizing the dilaton field -- the field whose vacuum value
determines the unified gauge coupling at the string scale. The
first example is based on the ``K\"ahler stabilization'' method
which assumes that nonperturbative corrections of string-theoretic
origin arise for the dilaton action~\cite{Sh90,BaDi94,Ca96}. Then
in the presence of one or more gaugino condensates in the hidden
sector the dilaton can be stabilized at weak coupling
($g_{\STR}^{2} = 1/2$) with a vanishing vacuum energy, provided
the parameters in the postulated nonperturbative correction are
chosen properly~\cite{BiGaWu97a,BiGaWu97b,BadeCo98}.

Such models give rise to dilaton-dominated supersymmetry breaking,
but the pattern of soft terms differs from the tree level examples
often studied. In particular, gaugino masses and A-terms are
suppressed relative to scalar masses by a loop factor with dilaton
contributions and contributions from the conformal anomaly
comparable in size. Examples of this scenario were presented
in~\cite{bench} and here we reproduce the case with condensing
group beta-function coefficient $b_+ = 9/16\pi^2$ with a gravitino
mass of $m_{3/2} = 4300 \GeV$.

\vspace{0.2in}
\noindent \textbf{Model G: Heterotic Strings with Racetrack
Stabilization.}

At the other extreme from the previous model is the case where
only the K\"ahler (compactification) moduli are involved in
transmitting the supersymmetry breaking to the observable sector.
This is a typical outcome in dilaton stabilization mechanisms that
employ the tree level K\"ahler potential for the dilaton, as in
the so-called ``racetrack method'' that uses multiple gaugino
condensates to stabilize the
dilaton~\cite{CaLaMuRo90,deCaMu93a,deCaMu93b}. Here the K\"ahler
moduli tend to be stabilized slightly away from their self-dual
points~\cite{FoIbLuQu90,CvFoIbLuQu91}.

When the observable sector matter fields all arise from the
untwisted sector then the entire soft supersymmetry breaking
Lagrangian arises only at the loop level. This was the case
studied in~\cite{bench}, where we here use the case with gravitino
mass of $m_{3/2} = 20\TeV$, Green-Schwarz counterterm coefficient
$\delta_{\GS} = -9$ and $\lang {\rm Re} \; T \rang = 1.23$.

\vspace{0.1in}
\noindent \textbf{Model H: Heterotic Strings with Strong
Coupling.}

This model explores the strong-coupling limit of heterotic $E_8
\times E_8$ string theory by studying the compactification of
11-dimensional M-theory on an $S_1/Z_2$
orbifold~\cite{HoWi96a,Wi96,HoWi96b}. The low-energy effective
Lagrangian for such a theory is characterized by deviations from
the weakly coupled case in both the gauge kinetic function and the
K\"ahler potential for the matter/moduli system~\cite{LuOvWa98a}.
These include the appearance of compactification moduli in the
gauge kinetic function as well as the appearance of the dilaton in
the kinetic function for the matter fields. These new
contributions are proportional to a constant that is computable
for a given compactification. The effect of these new terms is to
allow the F-terms for the compactification moduli to play a role
in determining soft gaugino masses while giving the dilaton F-term
a role in determining the scalar masses.

Using the soft terms expressions as given in~\cite{ChKiMu98}, we
have chosen a point in parameter space where both the dilaton and
the overall compactification modulus play equal roles in
supersymmetry breaking. The fundamental scale is taken to be
$\Lambda_{\UV} = \Lambda_{\GUT}$ with $g_{\STR}^{2} = 1/2$, but
with the size of the strong coupling correction such that this
occurs for $\lang {\rm Re}\; S \rang = 1.25$ instead of the usual
weakly coupled case of $\lang {\rm Re}\; S \rang = 2$.

\vspace{0.1in}
\noindent \textbf{Model I: Open Strings on Orientifolds with
Common $D_5$ Branes.}

In this example we consider a particular construction based on
Type IIB string theory with toroidal orientifold compactification
to four dimensions with $N=1$ supersymmetry~\cite{IbMuRi99}. For
maximum simplicity we will place the entire Standard Model gauge
group on a common set of $D_5$-branes whose world volumes are
parallel to one another. Up to new contributions to the gauge
kinetic functions from twisted moduli associated with the
blowing-up modes of the orbifold singularities, this construction
is very similar to the weakly coupled heterotic string at the tree
level. Neglecting these additional twisted moduli contributions to
gaugino masses is a good approximation in the limit where the
single modulus that determines the gauge coupling is the sole
participant in supersymmetry breaking~\cite{Be00}. In that limit
the soft terms are given by $M_{1/2} = -A_0 = \sqrt{3}m_0 =
\sqrt{3} m_{3/2}$, where $m_{3/2}$ is the gravitino mass.

This model has been studied at length in the
literature~\cite{BaLoMo93,BrIbMu94,CaLlMu96} and it represents a
special case of the mSUGRA paradigm. By couching it in the
language of $D_5$-branes, however, one is at liberty to set the
boundary condition (string) scale as a free parameter. The
superpartner spectrum of this model in the case where this scale
was taken to be $\Lambda_{\UV} = 1 \times 10^{11} \GeV$ was
studied in~\cite{AbAlIbKlQu00}. In order to achieve gauge coupling
unification it is necessary to add additional exotic matter states
to the theory at a low-energy scale. We follow a similar
prescription to that of~\cite{AbAlIbKlQu00} and add two
vector-like lepton doublets $(L,\oline{L})$ and three vector-like
singlets $(E,\oline{E})$ at the TeV scale with hypercharge equal
to their Standard Model equivalents. With this combination we find
the unification scale to be $\Lambda_{\UV} = 3 \times 10^{11}
\GeV$.

\vspace{0.1in}
\noindent \textbf{Model J: Open Strings on Orientifolds with
Intersecting $D_5$ Branes.}

In our next open string example we will allow the gauge groups of
the Standard Model to be localized on $D_5$-branes whose
world-volumes are not parallel. We have in mind the case where the
initial brane configuration is supersymmetric, with brane
world-volumes intersecting at right angles. To obtain a specific
model one must specify how fields and gauge groups are assigned to
the various 5-branes. In~\cite{BrEvKaLy99} the following scenario
was considered: assign the $SU(3)$ and $SU(2)$ gauge groups of the
Standard Model to separate 5-branes. For ease of memory we will
take $SU(3)$ to be on the $5_3$ brane and $SU(2)$ to be on the
$5_2$ brane. This means, for example, that the world volume of the
branes containing the $SU(3)$ gauge group have a world volume that
spans the third complex plane. Quark doublets must then be the
massless modes of strings that stretch between these branes.
Hypercharge must be represented by some linear combination of the
$U(1)$'s present on one or both branes. The decision of where to
put the hypercharge $U(1)$ is crucial to the string assignment of
the other Standard Model fields.

One possible configuration that allows for leading order Yukawa
couplings for all third generation particles is to place the
hypercharge factor solely on the $D_5$-branes containing the
$SU(2)$ sector. In this set-up all quark superfields arise from
the massless modes of stretched strings, while the lepton and
Higgs superfields arise from massless modes of strings which start
and end on the $5_2$ brane. To satisfy the string selection rules
for the Yukawa interactions, the Higgs superfields will then need
to have a K\"ahler potential with non-trivial dependence on the
modulus associated with the first complex plane. We then choose
the lepton singlet $E$ to have a K\"ahler metric which depends on
the modulus of the second complex plane while that of the lepton
doublet $L$ depends on the modulus of the first complex plane.

Assuming the perturbative (tree level) K\"ahler potential for all
moduli fields, and all K\"ahler moduli $T_1$, $T_2$ and $T_3$
participating in supersymmetry breaking equally, we then have
universal gaugino masses and trilinear couplings if the string
scale and GUT scale coincide. The scalar masses would be given by
$m_{H_u}^{2} = m_{H_d}^{2} = m_{E}^{2} =0$, $m_{Q}^{2} = m_{U}^{2}
= m_{D}^{2} =(1/2) m_{3/2}^{2}$, $m_{L}^{2} = m_{3/2}^{2}$ and
$M_{1/2} = -A_0 =  m_{3/2}$. At this point we may consider the
non-parallel $D_5$-brane construction as nothing more than a
motivation for studying a variant of the dilaton domination model
with a particular pattern of non-universal scalar masses at
$\Lambda_{\UV} = \Lambda_{\GUT}$.

\vspace{0.1in}
\noindent \textbf{Model K: Minimal $SU(5)$ with Gaugino
Mediation.}

The minimal $SU(5)$ GUT-based model assumes that the $SU(5)$ gauge
group is broken by the expectation value of a Higgs field in the
\textbf{24} representation of $SU(5)$ at a scale $\Lambda_{\GUT}
\simeq 2 \times 10^{16} \GeV$ to the Standard Model. For gaugino
mediation~\cite{KaKrSc00,ChLuNePo00} supersymmetry breaking is
assumed to not occur at this scale but rather at a higher scale
$M_c$ somewhere between the GUT scale and the Planck scale. As
such, the soft supersymmetry breaking terms at $M_c$ obey GUT
relations. The running of the soft Lagrangian between the two
scales produces variations between the soft scalar masses
associated with different representations under $SU(5)$, as well
as generations within the same representation class.

The soft terms of this minimal model are assumed to be based on
the notion that supersymmetry is broken in a hidden sector and is
communicated to the observable sector through gauge multiplets
which propagate in a higher-dimensional spacetime~\cite{ScSk00}.
The construction of the model is such that only gaugino masses
have non-zero soft masses at the leading order, with trilinear
A-terms, bilinear B-terms and scalar masses zero at the same
order. Collider signatures for the model described above were
considered by Baer et al.~\cite{BaBeKrTa02}. They chose to leave
the $\mu$ parameter and $B$ undetermined at the high-energy scale
so that they can be fit by the EWSB conditions (allowing
$\tan\beta$ to be a free parameter). We display results for their
example point that has the unified parameters $m_0 = 205.2 \GeV$
and $M_{1/2} = 400 \GeV$ with vanishing A-term and $\tan\beta
=35$, but with positive $\mu$ term. These parameters are run in
the GUT model from $M_c = 1 \times 10^{18} \GeV$ to $1.52 \times
10^{16} \GeV$ and are run in the general MSSM model thereafter.

\vspace{0.1in}
\noindent \textbf{Model L: GUT-Inspired Minimal $SO(10)$.}

The minimal $SO(10)$ GUT-based model of~\cite{BlDeRa02a,BlDeRa02b}
is based on the presumption that both gauge and Yukawa couplings
unify at some scale $\Lambda_{\GUT}$ into an $SO(10)$ framework.
Assuming supersymmetry to be broken at the GUT scale, the model is
defined by a universal gaugino mass $m_{1/2}$, a universal soft
scalar mass $m_{16}^{2}$ for the matter fields in the \textbf{16}
representation of SO(10), a universal soft scalar mass
$m_{10}^{2}$ for the Higgs doublets in the \textbf{10} of SO(10),
and a universal soft trilinear coupling $A_0$. To allow for proper
electroweak symmetry breaking at the large values of $\tan\beta$
typically required for Yukawa unification, the Higgs masses are
allowed to be split by some amount $\Delta m_{H}^{2}$, with the
up-type Higgs mass $m_{H_U}^{2}$ decreased by this amount while
the down-type Higgs mass $m_{H_D}^{2}$ is increased. We utilize a
set of soft parameters determined, through a global $\chi^2$
analysis, to produce consistent EWSB and gauge/Yukawa unification
at the scale $\Lambda_{\GUT} \simeq 3 \times 10^{16} \GeV$ given
known values of EW scale observables and fermion masses. The
values are consistent with the results of Tobe and
Wells~\cite{ToWe03}.

We consider one of the representative points of~\cite{BlDeRa02b}
with $m_{16} = 2000$, $m_{10} =2640$, $\Delta m_{H}^{2} = 0.13
m_{10}^{2}$, $A_0 = -3420$, $M_{1/2} = 350$ and $\mu=200$ in units
of GeV. The value of $\tan\beta$ is 52.5. Note that in this model
the boundary condition scale $\Lambda_{\GUT}$ is part of the
global fit, and is fixed to $3 \times 10^{16} \GeV$ for the
specific point we consider. While this is a GUT-inspired model, no
attempt is made to explain the Yukawa patterns of the first or
second generation fermions, though first and second generation
soft scalar masses are specified. Additionally, the specific model
of~\cite{BlDeRa02a,BlDeRa02b} did not specify any particular
neutrino sector, though with additional assumptions one could be
included quite easily.

\section{Discriminating Between Models with ``Observables''}
\label{sec:obs}

We have taken the mass spectrum generated by the~\modelnum~model
points of Section~\ref{sec:models} and estimated the inclusive
signatures of each case. By ``inclusive signature'' we are
referring to a signature that indicates the existence of physics
beyond the Standard Model, summed over all possible ways that such
a signature can arise. An inclusive signature must be an actual
physically observable quantity directly measured in experiments,
such as the excess above some background of jet events with
opposite-sign dileptons and missing transverse energy. It is
important to understand that experiments only measure rates and
kinematic distributions. Interpretation of results in terms of
superpartner masses such as the gluino mass are necessarily model
dependent and can be misleading. Essentially all soft Lagrangian
parameters such as gaugino masses, the $\mu$ parameter and
$\tan\beta$, etc. are unlikely to be directly measured.

The results of this estimation are included in the various rows of
Table~\ref{tbl:sig}. This is, of course, just a partial list for
purposes of illustration of the many measurable quantities that
can now, or in the near future, help us to learn about the
supersymmetric world. These quantities have been normalized in
such a way that a ``Y'' indicates the presence of an observation
that would clearly indicate new physics. More precise definitions
of the signatures and the criteria used for assigning a ``Y''
value can be found in the Appendix. Let us note that the Y/N
designation refers to the specific point in each model's parameter
space that was considered in Section~\ref{sec:models}. More
generally, a model with ``Y'' in a given observable may be thought
of as one which is {\em likely} to produce a signal throughout
most of its parameter space. In collider situations with limited
statistics (such as the Tevatron) or high backgrounds (such as the
LHC) these inclusive signatures will be the initial signals
observed. To go beyond them to soft Lagrangian parameters, or even
to identify which superpartners are being produced and measure
their masses, may require a long analysis and model-dependent
assumptions. Thus learning what we can from the inclusive
signatures alone could be essential.

The most important lesson to be learned from this table is that
even with the difficulties outlined above, such as the failure of
models to address vast amounts of data due to their incomplete
nature, {\em a handful of inclusive signatures may be sufficient
to distinguish classes of these models!} Strictly speaking, what
can be distinguished are the specific points within these models'
parameter spaces that we chose to study. But to the extent that
these points are truly representative then this is not merely an
artifact of the points in parameter space that were chosen, but a
property of the very real differences between the structure of
these models. We expect that this is likely to be a robust feature
of even wider arrays of models. Note that no specific soft
Lagrangian parameters need to be measured -- not even $\tan\beta$,
whose value we have suppressed on purpose. The measurements
envisioned in Table~\ref{tbl:sig} are in no sense precision
measurements but merely observations/non-observations. The data as
seen by the experimentalist is similar to the columns we present
here -- a sequence of observations/non-observations with no
``model name'' to associate with the data or to assist in its
interpretation. An important point is that experimenters can
compare total event rates with those expected from the Standard
Model and need not reduce samples with cuts.

%
\begin{table}[p]
\begin{center}
\begin{footnotesize}
\begin{tabular}{|p{0.1cm}l|c|c|c|c|c|c|c|c|c|c|c|c|}\hline
\multicolumn{2}{|l|}{Inclusive Signature} & A & B & C & D & E & F
& G & H & I & J & K & L
\\ \hline \hline
\multicolumn{2}{|l|}{Collider} &  &  &  &  &  &  &  &  &  &  & &
\\
 & Large $\notE$ & Y & Y & Y & Y & Y & Y & Y & Y & Y & Y & Y & Y \\
 & Prompt $\gamma$ & N & N & Y & N & N & N & N & N & N & N & N & N \\
 & Isolated $\pi^{\pm}$ & N & N & N & N & N & N & Y & N & N & N & N & N \\
 & Trilepton  & Y & Y & Y & N & N & Y & N & Y & Y & Y & Y & Y \\
 & SS dilepton  & Y & Y & Y & N & N & Y & N & Y & Y & Y & Y & Y \\
 & OS dilepton  & Y & N & Y & N & N & Y & N & Y & Y & Y & Y & N \\
 & $\tau$ rich  & N & N & N & N & N & N & N & N & N & N & N & N \\
 & $b$ rich  & N & N & N & N & N & N & N & N & N & N & N & N \\
 & Long-lived (N)LSP  & Y & Y & Y & Y & Y & Y & Y & Y & Y & Y & Y & Y \\
\hline
\multicolumn{2}{|l|}{Non-SM Flavor and CP} &  &  &  &  &  &  &  &
& & & & \\
 & $g_{\mu} -2$  & Y & N & Y & N & N & N & N & Y & N & N & Y & N \\
 & $B_s \to \mu^+ \mu^- $  & N & N & N & N & N & N & N & N & N & N & N & Y \\
 & $B \to X_s \gamma$ & \checkmark & \checkmark & \checkmark &
 \checkmark & \checkmark & \checkmark & \checkmark & \checkmark
 & \checkmark & \checkmark & \checkmark & \checkmark \\
 & $A_{\CP}\(B\to s \gamma\)$  & $\SM$ & $\SM$ & $\SM$ & $\SM$ & $\SM$
 & $\SM$ & $\SM$ & $\SM$ & $\SM$ & $\SM$ & $\SM$ & $\SM$ \\
 & $A_{\CP}\(B\to \phi K_S\)$ & $\SM$ & $\SM$ & $\SM$ & $\SM$ & $\SM$
 & $\SM$ & $\SM$ & $\SM$ & $\SM$ & $\SM$ & $\SM$ & $\SM$ \\
 & $K-\oline{K}$ mixing & $\SM$ & $\SM$ & $\SM$ & $\SM$ & $\SM$
 & $\SM$ & $\SM$ & $\SM$ & $\SM$ & $\SM$ & $\SM$ & $\SM$ \\
 & $\epsilon'/\epsilon$ & $\SM$ & $\SM$ & $\SM$ & $\SM$ & $\SM$
 & $\SM$ & $\SM$ & $\SM$ & $\SM$ & $\SM$ & $\SM$ & $\SM$ \\
 & $\mu \to e \gamma$ & $\SM$ & $\SM$ & $\SM$ & $\SM$ & $\SM$
 & $\SM$ & $\SM$ & $\SM$ & $\SM$ & $\SM$ & $\SM$ & $\SM$ \\
 & $(2\beta)_{0\nu}$  &  &  &  &  &  &  &  &  &  &  &  & \\
 & eEDM, qEDM  & $\SM$ & $\SM$ & $\SM$ & $\SM$ & $\SM$
 & $\SM$ & $\SM$ & $\SM$ & $\SM$ & $\SM$ & $\SM$ & $\SM$ \\
%
%
\hline
\multicolumn{2}{|l|}{Cosmology} &  &  &  &  &  &  &  &  &  &  & &
\\
 & Direct WIMP detection  & N & N & N & N & N & N & N & N & N
 & N & N & ${\rm Y}^{*}$ \\
 & Space-based signals ($e^{+}, \bar{p}, \; \gamma$)
 & N & N & N & ${\rm Y}^{*}$ & ${\rm Y}^{*}$ & N & N & N & N & N & N & ${\rm Y}^{*}$ \\
 & Neutrinos from LSP annihilation
 & N & N & N & N & N & N & N & N & N & N & N & ${\rm Y}^{*}$ \\
 & Detectable axion  &  &  &  &  &  &  &  &  &  &  &  & \\
 & Baryon asymmetry  &  &  &  &  &  &  &  &  &  &  &  & \\
\hline
\multicolumn{2}{|l|}{EWSB Sector} &  &  &  & & & & & & & & & \\
 & $M_Z$  &  &  &  &  &  &  &  &  &  &  &  & \\
 & EW precision data & \checkmark & \checkmark &
 \checkmark & \checkmark & \checkmark & \checkmark & \checkmark &
 \checkmark & \checkmark & \checkmark & \checkmark & \checkmark \\
\hline
\multicolumn{2}{|l|}{Unification and Extended Sectors} &  &  &  &
& & & & & & & & \\
 & Unified $\alpha_{\GUT} (\Lambda_{\UV})$ & \checkmark & \checkmark &
 \checkmark & \checkmark & \checkmark & \checkmark & \checkmark &
 \checkmark & \checkmark & \checkmark & \checkmark & \checkmark \\
 & $\alpha_s (M_Z) = 0.118$  &  &  &  &  &  &  &  &  &  &  &  &  \\
%
%
 & TeV-scale exotic particles  & N & N & N & N & N & N & N & N & Y & N & N & N \\
 & Proton decay  &  &  &  &  &  &  &  &  &  &  &  &  \\
\hline \hline
\end{tabular}
\caption{\footnotesize \label{tbl:sig} Inclusive signature list
for the~\modelnum~models. The general meaning of each of the
signatures listed in the first column is given in the Appendix.
Collider signatures are normalized to an LHC luminosity of $10
{\rm fb}^{-1}$ for one year of running with a $5 \sigma$ discovery
threshold. We list a number of possible inclusive signatures to
show what is possible, but at the present time there are few
models that include all of the relevant physics. For observables
that are basically ignored by the models we either leave the row
blank, state that it gives the SM result in a trivial way, or use
a \checkmark to indicate a non-trivial consistency check.}
\end{footnotesize}
\end{center}
\end{table}

We break the listing into five categories: collider signatures,
signatures of new physics in the flavor and CP violating arenas,
cosmological signatures, the electroweak symmetry-breaking sector
and signatures of models that go beyond the MSSM in content. Not
surprisingly, all models are capable of making predictions in the
collider arena and in most of the cosmological section since they
all (at a minimum) predict the superpartner spectrum. But many
rows are blank or give only the Standard Model result ($\SM$) in a
trivial way. This is a reflection of the fact that none of these
models address the strong CP problem (axions), have a neutrino
sector ($(2\beta)_{0\nu}$), have phases (EDMs and CP asymmetries),
or postulate specific Yukawa textures. All take R-parity
conservation as a given fact and none address higher order
operators, so none can make concrete predictions about proton
decay (though some may allow it). All these models can say
something about rare decays that could yield signals of new
physics only in the limit where a signal can occur in the minimal
flavor violation paradigm (such as the process $B_s \to \mu^+
\mu^- $ or the muon anomalous magnetic moment\footnote{We here
take an agnostic view and assume for the purpose of this letter
that the measurement of $(g_{\mu}-2)$ does not yet represent a
signal for new physics}), since all take the CKM matrix as inputs.

Still other observable quantities which have already been measured
(such as the Z-boson mass and the strong coupling constant
$\alpha_s(M_Z)$) are taken as input quantities by all of these
models. Thus the models are trivially consistent with these
observations and do not {\em explain} these measured values.
Indeed, this lack of explanation tends to reflect itself in large
fine-tunings in, for example, the EWSB sector. To the extent that
all of these models are designed to ensure gauge coupling
unification, and assume all flavor structure is contained within
the Standard Model CKM matrix, they can (in some limited sense)
make predictions for the values of quantities such as the
branching ratio ${\rm Br}(b \to s \gamma)$ and various electroweak
precision variables. That they are consistent with the measured
results (i.e. consistent with the hypothesis of no significant
SUSY contributions) is merely a reflection of these starting
assumptions. We have indicated this consistency with a check mark
in Table~\ref{tbl:sig}.

Ultimately a complete model would provide meaningful predictions
in the form of ``Y'' or ``N'' for each of these observables, or a
``\checkmark'' that truly reflects a non-trivial consistency check
on the theory. For example, most existing models would naively
predict $M_Z$ to be an order of magnitude larger than it is if
this mass were not already known. In each of the~\modelnum~models
we consider this constraint is satisfied by taking $M_Z$ as an
input rather than an output and assuming a fixed value of $\mu$ at
the high scale to ensure this outcome. Without an explicit
$\mu$-term generating mechanism this cannot be seen as a
``prediction'' for $M_Z$. As more observables are measured various
rows in the table which currently display ``Y/N'' predictions
become measured constraints. Models incapable of satisfying these
constraints must, of course, then be modified or discarded.

Finally, there is yet another set of measured quantities that may
be of use in determining the right supersymmetric Standard Model.
We call these ``derived observables'' in that they are based on
actually measured quantities, but the interpretation of that
measurement can only be done in the context of a specific theory.
For example, the WMAP experiment achieves a precise measurement of
the non-baryonic cold dark matter density by fitting the observed
CMB power spectrum to a number of input
variables~\cite{Spergel:2003cb}. Thus, this quantity is
measurable. If we wish to identify this observed matter component
with a supersymmetric particle such as the lightest neutralino,
then the necessary ingredients must be incorporated into a
supersymmetric model: R-parity conservation, a relic production
mechanism of thermal or non-thermal nature, and so on. Then this
measurement becomes a constraint on the remaining parameter space
(such as soft masses and $\tan\beta$) of this expanded model. But
because these quantities either do not exist or are not calculable
within the Standard Model, without a model it is not even clear
what the WMAP observation is really measuring. Thus models that
have all the requirements to make definite predictions about cold
dark matter might be ruled out by this measurement -- those that
don't are unaffected.

Or take the case of coupling unification -- both gauge and Yukawa
couplings. Strictly speaking this is a high-energy prediction of
some models that is not directly testable. But once a model
postulates the complete matter and gauge content of a theory, as
well as a fundamental scale, then the observed gauge couplings are
sufficient to ``measure'' the existence of a unified gauge
coupling at some higher scale. We have included this ``derived
observable'' in the Unification section of Table~\ref{tbl:sig}
since each of these models postulates the necessary assumptions to
make such a derived quantity meaningful. So too the measured
fermion masses, when coupled with an eventual measurement of
$\tan\beta$ produces a ``measurement'' of Yukawa unification in
the context of a complete model. Without such a complete model,
these UV scale relations cease to have any meaning as observables.

Another example might be the degree of fine-tuning involved in
satisfying a constraint. Once a quantity is measured, the tuning
required to achieve this outcome is a legitimate observable to
consider. While not directly measurable experimentally, once
defined such tunings become a quantifiable derived observable that
distinguishes between models. The discriminatory power of these
derived observables is of help to the theorist in seeking ways to
interpret the data. The mass of the Z-boson has been measured;
hence the tuning on $M_Z$, however it is defined, becomes a
distinguishing (theoretical) characteristic of models. Even the
observed acceleration in the supernovas recession rate and the
observable quantities associated with the cosmic microwave
background radiation can become potential derived observables. We
have not included such signatures in Table~\ref{tbl:sig} but once
a model identifies the inflaton with a field of the supersymmetric
Standard Model, for example, then these parameters become fair
constraints that such a model must satisfy.

\section{Phenomenological Analysis}
\label{sec:pheno}

\subsection{Discovery and Interpretation of Superpartners}
\label{sec:disc}

While we may be lucky and find a superpartner signal at the
Tevatron in spite of its poor performance, signals will surely
appear at the LHC if low scale supersymmetry is present in nature.
So let us present our analysis for LHC.\footnote{The approach we
describe will be even more essential if superpartners are
discovered at the Tevatron.} Suppose supersymmetry-like excesses
of events begin to be reported at LHC. What can be learned after
the excitement of the raw discovery, given that there will be too
few observables to deduce the underlying Lagrangian? To study this
question we first simulate a possible signal, and then examine how
far we can go in recovering the underlying physics.

The LHC experiments will measure production cross sections times
branching ratios, and some kinematic distributions that will
provide information about combinations of the eigenvalues of
superpartner mass matrices. We assume $10 {\rm fb}^{-1}$
integrated luminosity, i.e. one year at $10^{33} {\rm cm}^{-2}
{\rm sec} ^{-1}$, and consider the typical channels that will be
studied. We further assume all channels involve missing transverse
energy larger than $100 \GeV$ and at least two jets, each with
transverse energy above $100\GeV$. The excesses that are
``discovered'' differ in their leptonic properties (isolated
energetic leptons). For the SM about 100,000 events are found with
no leptons, 13,000 with one lepton, 7,000 with two opposite sign
leptons, 20 with two same sign leptons, and 60 trileptons, all
with large missing transverse energy. One model~(Model F) gives
respectively 31,700; 7,300; 2,000; 504; and 204 events for these
channels in excess of the SM. Thus all channels show very
statistically significant signals if we just use $\sqrt{N_{\SM}}$
as an approximate standard deviation. In addition, the peak of the
so-called $m_{\rm eff}$ distribution (basically the sum of all
transverse and missing energy) is at $838\GeV$ for Model F. Some
other kinematic distributions can be measured, such as the end
points of the lepton $p_{T}$ distributions. There may also be
information about the Higgs sector; for simplicity we won't
include that here. Table~\ref{tbl:num_events} shows the excesses
for some of the other models in addition to Model F, as well as
the Standard Model baseline.

\begin{table}[ht]{}
\begin{center}
\begin{tabular}{|c|c||c|c|c|c|c|c|c|c|c|} \hline
Channel & SM & A & B & C & F & H & I & J & K & L \\ \hline \hline
Jets ($\times 10^3$) & 100.0 & 59.5 & 0.7& 4.2 & 31.7& 6.6 & 5.0 &
7.2 & 7.0 & 1.1 \\
$1\ell$ ($\times 10^3$) & 13.0 & 17.1 & 0.5 & 1.8 & 7.3 & 1.7 &
1.8 & 1.8 & 1.9 & 0.5 \\
OS ($\times 10^3$) & 7.0 & 5.7 & 0.2 & 1.1 & 2.0 & 0.6 & 0.8 & 0.8
& 0.6 & 0.2 \\
SS & 20 & 1332 & 99 & 277 & 504& 160 & 252 & 155 & 197 & 90 \\
$3\ell$ & 60 & 737 & 97 & 310 & 204 & 77 & 137 & 111 & 71 & 82 \\
$m_{\rm eff}^{\rm peak}$ (GeV) & - & 812 & 1140 & 1310 & 838 &
1210 & 1210 & 1340 & 1290 & 1210 \\ \hline
\end{tabular}
\end{center}
\caption{\label{tbl:num_events} \footnotesize {\bf Number of
events in excess of the Standard Model prediction for different
signatures.} For each channel the Standard Model baseline is given
in the first column. Subsequent columns give the excess beyond
this baseline for selected models from Section~\ref{sec:models}. }
\end{table}

What can be learned from this information? From the event rates
and distributions information can be obtained on the gluino mass
and some squark, slepton, chargino and neutralino masses. The
masses themselves cannot be measured accurately since there are
not only significant experimental errors, but also considerable
model dependence. Perhaps the lightest superpartner (LSP) mass can
be measured at the 10-20\% level. Remember that none of these
masses correspond to Lagrangian parameters since they are
radiatively corrected mass eigenstates, so we cannot determine the
elements of the mass matrices (i.e. $\tan \beta$, the $\mu $
parameter magnitude and phase, the gaugino mass parameters
$M_{1}$, $M_{2}$, $M_{3}$, etc). Even though we can't determine
these parameters, can we nevertheless extract information about
the underlying theory?

Experimental analyses are (necessarily) done independently channel
by channel. Measurement of the excess in any given channel will
tell us little about the underlying theory since many sets of
parameters in a given model, and many kinds of models, will give
about the right number. What if we combine several channels? One
can discuss this issue at several levels. First, suppose one knew
or assumed a given basic model that contained several Lagrangian
parameters. Can one then from a few actual experimental
observables (that we call inclusive signatures) determine the
basic parameters such as $\tan\beta $, $\mu$, gaugino masses,
phases, etc.?  Note that any such determination would necessarily
have an associated model dependence because {\em a priori} \, some
other model might describe the data as well. We find that using
the procedures we describe below leads to significant success
here.

Second and more important, but more uncertain: could one, from a
set of inclusive signatures, actually determine that some
underlying theories (including the criteria which describe
supersymmetry breaking itself) were significantly favored relative
to others? In one sense this is not so hard, since some types of
signatures are obviously special to some types of theories. For
example, prompt photons only occur in gravity mediated theories if
the LSP is Higgsino-like, and then only in certain types of
events, while in gauge mediated theories every event has two
prompt photons (unless the LSP is long-lived which can be tested
for in other ways). But unless a lucky signature such as this
occurs, it will be much more difficult to gain information about
supersymmetry breaking or the high scale effective Lagrangian, let
alone the corner of M-theory that is being seen, by normal
methods. Consequently we have tried to develop an algorithmic
approach that may lead to insights into the underlying theory even
from inclusive signatures, and even when crucial quantities such
as $\tan\beta$ have not yet been measured. We study these issues
using models and simulated data in the following section.

To avoid misunderstanding we emphasize that the approach we
describe is to help guide theorists toward what kind(s) of
underlying theories and supersymmetry breaking to focus on. It is
not meant to prove one theory is right, but to learn better what
additional observables may provide particular sensitivity and what
aspects of the theory need improved study and calculation. The
method, a ``global fit,'' is in principle valid because all
approaches, whatever the stringy connection and however
supersymmetry is broken, connect to the observable world via the
soft supersymmetry breaking Lagrangian.

\subsection{Global theoretical fits}
\label{sec:chi}

We study here an approach that we think may become a powerful
technique to achieve the goals discussed above; that is, to relate
incomplete data to the underlying theory. In essence, what we
require is simultaneous consistency with several pieces of
information ({\em e.g.} collider data, rare decays, relic
neutralino interaction rates, and perhaps more as described
below), and provide a quantitative measure to compare models and
classes of models.

The idea of using likelihood techniques to identify favored areas
of a model's parameter space is certainly not new.\footnote{In
fact, as this work was nearing completion a likelihood analysis of
the mSUGRA parameter space appeared in~\cite{Ellis}.} The classic
example is the global fit of the parameters of the Standard Model
to electroweak precision measurements~\cite{ErLa95}. However, the
notion that such an approach can be used to {\em distinguish}
classes of models has not to our knowledge been pursued. Previous
uses of this approach, which we call a $\chi_T^2$ (or theory
$\chi^2$) analysis, tend to focus on only one model and often
include ``observations'' that we have classified in
Section~\ref{sec:obs} as derived observables such as Yukawa
unification or the thermal relic density of neutralinos.

Rather than using a likelihood technique to find the most-favored
point of a given parameter space, we propose using the $\chi_T^2$
technique to find the most-favored {\em model} in the space of
theories. Let us see how such a procedure might work in the case
of the~\modelnum~models of Section~\ref{sec:models}. Consider the
minimal supergravity model with the following typical parameter
set
\begin{equation}
\tan\beta=10 \quad m_{1/2}=380 \quad m_0=500 \quad A_0 =0 \quad
{\rm sgn}(\mu)>0.
\label{sugra} \end{equation}
The resulting low-energy soft Lagrangian is obtained through RG
evolution of these parameters using {\tt SuSpect}~\cite{SUSPECT}.
At the electroweak scale these values can be passed to {\tt
PYTHIA}~\cite{Sjostrand:2000wi} and an LHC collider simulation
performed. We use the results of this simulation to calculate the
number of events for the five inclusive signatures listed in
Table~\ref{tbl:num_events} for a given luminosity.

Imagine the situation after the first year of data collection at
the LHC. Experimental results for any excess {\em above} the
Standard Model prediction, in each of the channels in
Table~\ref{tbl:num_events}, will be reported. Let us denote this
excess by the vector
\begin{equation}
\vec{a}^{\rm exp}=\{a_i^{\rm exp}\} \quad i=1,2,\cdots,5 ,
\label{LHCresults} \end{equation}
and the expected Standard Model background by
\begin{equation}
\vec{a}^{\SM}=\{a_i^{\SM}\} \quad i=1,2,\cdots,5 .
\label{background} \end{equation}
An estimation of the sizes of these background values for the LHC
can be found in~\cite{Baer:1995nq,Baer:1995va}.

Can we use the ``experimental'' result of~(\ref{LHCresults}) to
reconstruct the mSUGRA point of~(\ref{sugra})? We imagine dividing
the parameter space of any model into a coarse grid with
coordinate given by the parameter vector $\vec{x}$. For example,
in the minimal supergravity model we would have
\begin{equation}
\vec{x}=\{m_{1/2},m_0,\tan\beta,A_0,{\rm sgn}(\mu)\} ,
\label{xsugra} \end{equation}
while in the dilaton-dominated model $\vec{x} = \{ m_{3/2}, b_+,
\tan\beta\}$, and so forth. For each point labeled by $\vec{x}$ we
will simulate some number of events (for illustrative purposes for
this paper 30,000 events) within the framework of a particular
model and compute the theoretical prediction $\vec{a}^{\rm
th}=\{a_i^{\rm th}\}$. We then construct the quantity
\begin{equation}
\chi^2_T = \sum_{i=1}^{5}\frac{|a^{\rm th}_i(\vec{x})-a^{\rm
exp}_i|^2}{\sigma_i^2}
\label{chi2} \end{equation}
where $\sigma_i^2$ is an experimental uncertainty of each value
$a^{\rm exp}_i$ that we assign. For a Poisson distribution,
$\sigma_i=\sqrt{a^{\rm exp}_i+a^{\SM}_i}$. Note that with this
definition the contribution to $\chi^2_T$ from collider
observables is proportional to the luminosity, so care is needed
in interpreting $\chi^2_T$ and in choosing relative weights of
collider and non-collider observables.

At this point questions could be raised about statistical
measures, independent variables, etc. We think that is not a
productive set of issues -- at least not at this stage. In
principle all of the models we study, to the extent that they are
complete, specify a value for the 105 parameters of the MSSM. In
most of the models we treat here the vast majority of these
parameters are either zero or related in some simple manner.
Nevertheless, one can imagine the space of these theories as being
the space of the MSSM parameter set so that our theoretical
measure is finding the best set of parameters for {\em one} model
-- the MSSM -- but that best set may only be consistent with one
class of MSSM paradigms represented by a particular type of
theory. Theories where many of the parameters vanish or are
strongly related tend to yield ``$\SM$'' in Table~\ref{tbl:sig}.
But this merely indicates that some parameters in the theory (say,
for example, the flavor structure of the soft
supersymmetry-breaking trilinear couplings) play no role in
determining the $\chi^2_T$ for that theory. In other cases where
the theory makes a non-trivial prediction the quantities which
receive ``Y/N'' predictions or ``$\checkmark$'' are included in
the fit. In some sense the main difference, then, between this
$\chi^2_T$ variable and the standard one used by experimentalists
is that we seek to include only inclusive signatures. This measure
can help us distinguish between models that appear, at first
sight, to be extremely similar in nature. Two models that look
similar in terms of there Y/N predictions in Table~\ref{tbl:sig}
may yield vastly different $\chi^2_T$ values in the fit,
suggesting a differentiation not apparent before. This power is
augmented when sensitivity, or fine-tuning measures are included
in the analysis. The approach here should be treated as providing
guidance and should not be used for conclusions such as ``the
heterotic string theory on an orbifold is $3.2 \sigma$ better at
fitting the data than the Type I theory...''.

\begin{table}[ht]
\begin{center}
\begin{tabular}{|l||l|l|l|l|l|l|l|l|l|l|} \hline
 & \multicolumn{10}{|c|}{$m_{1/2}$ (GeV)} \\ \hline
$m_{0}$ (GeV) & 300 & 320 & 340 & 360 & 380 & 400 & 420 & 440 &
460 & 480 \\ \hline \hline
100 & 1097 & 471 & 175 & 60.6 & 16.9 & 3.6 & 2.5 & 6.4 & 12.7 &
19.4 \\ \hline
200 & 792 & 363 & 159 & 63.3 & 20.1 & 6.3 & 7.2 & 15.1 & 23.5 &
34.7 \\ \hline
300 & 528 & 244 & 94.9 & 36.6 & 15.5 & 10.9 & 14.2 & 23.8 & 33.5 &
43.6 \\ \hline
400 & 445 & 166 & 53.7 & 16.3 & 5.7 & 8.2 & 14.7 & 27.8 & 37.1 &
46.0 \\ \hline
500 & 427 & 207 & 37.2 & 9.3 & \textbf{1.7} & 7.6 & 16.5 & 28.7 &
39.7 & 49.4 \\ \hline
600 & 248 & 668 & 70.6 & 20.6 & 5.5 & 10.0 & 20.8 & 31.0 & 42.2 &
51.7 \\ \hline
700 & 197 & 255 & 136 & 35.6 & 16.2 & 16.2 & 21.9 & 33.7 & 44.9 &
54.9 \\ \hline
800 & 178 & 214 & 57.0 & 51.2 & 18.7 & 23.3 & 28.3 & 36.9 & 45.5 &
55.2 \\ \hline
900 & 133 & 34.4 & 27.8 & 27.3 & 29.1 & 29.8 & 34.8 & 41.4 & 49.5
& 56.2 \\ \hline
1000& 110 & 23.8 & 20.0 & 26.8 & 29.0 & 37.1 & 41.6 & 48.8 & 54.8
& 60.7 \\ \hline
\end{tabular}
\end{center}
\caption{\label{tbl:sugra_chi_A} \footnotesize \textbf{Values of
$\chi^2_T$ for the base mSUGRA model given by~(\ref{sugra}).}}
\end{table}

With these caveats in mind, we can now perform for each model a
global inclusive fit to the theory by seeking the minimum
$\chi^2_T$ by varying $\vec{x}$. Call this minimum $\chi^2_T$ as
$\(\chi^2_T\)_{m}$. By comparing $\(\chi^2_T\)_{m}$ for different
theories we can crudely identify the theory with the smallest
value of $\(\chi^2_T\)_{m}$ as the most promising candidate.
Furthermore, theories that yield $\(\chi^2_T\)_{m} \gg d$, where
$d$ is the number of degrees of freedom (roughly the dimension of
the vector $\vec{a}^{\rm exp}$ minus the dimension of $\vec{x}$)
should be strongly disfavored. For the analysis we present below
the five collider variables of Table~\ref{tbl:num_events} were
included in all cases. In addition to the hypothesized LHC data we
can also include other experimental results in the $\chi^2_T$ in a
similar manner. In particular we also include experimental
measurements of ${\rm Br}(b \to s \gamma)$ and $g_{\mu} -2$ with
the following values: $a_{\mu}^{\SUSY} = (20 \pm 20) \times
10^{-10}$ and ${\rm Br}(b \to s \gamma) = (3.25 \pm 0.37) \times
10^{-4}$. Given that our purpose here is merely to illustrate the
method, the actual values chosen are not important.

\begin{table}[t]
\begin{center}
\begin{tabular}{|l||l|l|l|l|l|l|l|l|l|l|} \hline
 & \multicolumn{10}{|c|}{$m_{1/2}$ (GeV)} \\ \hline
$m_{0}$ (GeV) & 300 & 320 & 340 & 360 & 380 & 400 & 420 & 440 &
460 & 480 \\ \hline \hline
100 & $\NA$ & $\NA$ & $\NA$ & $\NA$ & $\NA$ & $\NA$ & $\NA$ &
$\NA$ & $\NA$ & $\NA$ \\ \hline
200 & 616 & 275 & 119 & 59.2 & $\NA$ & $\NA$ & $\NA$ & $\NA$ &
$\NA$ & $\NA$ \\ \hline
300 & 617 & 271 & 115 & 51.4 & 16.2 & 4.8 & 7.3 & 15.1 & 26.5 &
35.9 \\ \hline
400 & 442 & 192 & 59.5 & 21.1 & 8.2 & 8.5 & 16.7 & 23.5 & 32.9 &
39.1 \\ \hline
500 & 383 & 135 & 35.6 & 7.4 & 2.7 & 7.8 & 17.3 & 27.5 & 37.7 &
48.1 \\ \hline
600 & 203 & 139 & 26.6 & 5.7 & \textbf{2.4} & 10.1 & 21.7 & 32.6 &
42.2 & 50.7 \\ \hline
700 & 141 & 80.6 & 32.0 & 7.9 & 7.9 & 13.4 & 24.6 & 34.1 & 45.6 &
55.6 \\ \hline
800 & 136 & 95.5 & 34.8 & 22.3 & 34.7 & 20.3 & 28.3 & 37.4 & 45.8
& 55.8
\\ \hline
900 & 90.5 & 76.1 & 20.7 & 24.7 & 29.7 & 30.1 & 35.5 & 42.9 & 51.2
& 58.2 \\ \hline
1000& 80.6 & 54.3 & 27.4 & 21.5 & 27.8 & 34.6 & 39.7 & 44.2 & 55.5
& 62.2 \\ \hline
\end{tabular}
\end{center}
\caption{\label{tbl:sugra_chi_B} \footnotesize \textbf{Values of
$\chi^2_T$ for the base mSUGRA model given by~(\ref{sugra}) but
with $\tan\beta=40$.} Entries marked $\NA$ have a stau LSP.}
\end{table}

Let us begin with the minimal supergravity model. The value of
$\chi_T^2$ for each point in our coarse grid is given for $\mu >
0$, $A_0 = 0$ and $\tan\beta = 10$ in Table~\ref{tbl:sugra_chi_A}.
The value of $\(\chi^2_T\)_m = 1.7$ does indeed fall at the value
input from~(\ref{sugra}). But in Table~\ref{tbl:sugra_chi_A} we
assumed the correct values of $\tan\beta$ and ${\rm sgn}(\mu)$. In
Tables~\ref{tbl:sugra_chi_B} and~\ref{tbl:sugra_chi_C} we give the
same region of the $(m_{1/2}, m_0)$ plane but change the value of
$\tan\beta$ to $\tan\beta=40$ in Table~\ref{tbl:sugra_chi_B} and
the sign of $\mu$ in Table~\ref{tbl:sugra_chi_C}. In the former
case the location of $\(\chi^2_T\)_m$ is displaced slightly while
its value differs only slightly from the input case, so these
inclusive signatures are not very sensitive to $\tan\beta$. In the
latter case the location of $\(\chi^2_T\)_m$ is unchanged but the
value has increased somewhat more. When both values are altered,
as in Table~\ref{tbl:sugra_chi_D}, the location of the minimum
shifts dramatically and the value of $\(\chi^2_T\)_m$ is much
larger.

\begin{table}[tp]
\begin{center}
\begin{tabular}{|l||l|l|l|l|l|l|l|l|l|l|} \hline
 & \multicolumn{10}{|c|}{$m_{1/2}$ (GeV)} \\ \hline
$m_{0}$ (GeV) & 300 & 320 & 340 & 360 & 380 & 400 & 420 & 440 &
460 & 480 \\ \hline \hline
100 & 689 & 326 & 151 & 89.8 & 64.1 & 40.8 & 31.6 & 28.6 & 28.8 &
30.7 \\ \hline
200 & 793 & 380 & 174 & 73.5 & 29.6 & 16.1 & 13.1 & 21.3 & 30.9 &
38.6 \\ \hline
300 & 512 & 232 & 102 & 47.9 & 24.3 & 17.5 & 23.3 & 30.5 & 37.6 &
45.3 \\ \hline
400 & 408 & 163 & 64.5 & 22.9 & 11.3 & 13.9 & 21.4 & 31.1 & 42.4 &
49.0 \\ \hline
500 & 457 & 144 & 46.1 & 14.9 & \textbf{7.0} & 11.2 & 21.7 & 32.7
& 42.2 & 52.0 \\ \hline
600 & 233 & 261 & 80.2 & 29.4 & 9.7 & 14.1 & 23.1 & 33.8 & 45.1 &
53.5 \\ \hline
700 & 188 & 176 & 144 & 47.8 & 22.3 & 19.0 & 25.4 & 35.8 & 45.3 &
55.6 \\ \hline
800 & 146.8 & 51.4 & 63.7 & 63.7 & 68.3 & 24.9 & 30.6 & 38.2 &
46.3 & 56.0 \\ \hline
900 & 106 & 28.2 & 44.8 & 35.9 & 35.0 & 38.7 & 36.5 & 42.8 & 49.8
& 57.8 \\ \hline
1000 & 79.7 & 19.8 & 24.6 & 26.5 & 30.9 & 37.8 & 42.9 & 51.6 &
56.2 & 62.3 \\ \hline
\end{tabular}
\end{center}
\caption{\label{tbl:sugra_chi_C} \footnotesize \textbf{Values of
$\chi^2_T$ for the base mSUGRA model given by~(\ref{sugra}) but
with $\tan\beta=10$ and $\mu <0$.}}
\end{table}

\begin{table}[h]
\begin{center}
\begin{tabular}{|l||l|l|l|l|l|l|l|l|l|l|} \hline
 & \multicolumn{10}{|c|}{$m_{1/2}$ (GeV)} \\ \hline
$m_{0}$ (GeV) & 300 & 320 & 340 & 360 & 380 & 400 & 420 & 440 &
460 & 480 \\ \hline \hline
100 & $\NA$ & $\NA$ & $\NA$ & $\NA$ & $\NA$ & $\NA$ & $\NA$ &
$\NA$ & $\NA$ & $\NA$ \\ \hline
200 & 1316 & 800 & 560 & 427 & 355 & 313 & 286 & 261 & 246 & $\NA$
\\ \hline
300 & 1043 & 633 & 433 & 326 & 269 & 234 & 214 & 204 & 198 & 193
\\ \hline
400 & 684 & 421 & 280 & 221 & 190 & 177 & 171 & 170 & 170 & 167 \\
\hline
500 & 489 & 262 & 171 & 138 & 126 & 124 & 128 & 133 & 138 & 143 \\
\hline
600 & 297 & 204 & 114 & 86.4 & 81.6 & 86.8 & 95.3 & 104 & 110 &
118 \\ \hline
700 & 187 & 107 & 77.0 & 58.1 & 56.7 & 62.1 & 70.9 & 81.7 & 91.8 &
99.9 \\ \hline
800 & 141 & 85.2 & 52.1 & 48.9 & 55.0 & 50.6 & 58.2 & 68.1 & 78.1
& 85.2 \\ \hline
900 & 107 & 95.8 & 58.8 & 44.7 & 46.6 & 47.5 & 55.6 & 62.3 & 69.9
& 77.6 \\ \hline
1000 & 81.1 & 74.0 & \textbf{31.2} & 37.2 & 41.2 & 47.9 & 52.6 &
57.4 & 68.5 & 74.5 \\ \hline
\end{tabular}
\end{center}
\caption{\label{tbl:sugra_chi_D} \footnotesize \textbf{Values of
$\chi^2_T$ for the base mSUGRA model given by~(\ref{sugra}) but
with $\tan\beta=40$ and $\mu <0$.} Entries marked $\NA$ have a
stau LSP.}
\end{table}

Whether such differences are meaningful would require a more
careful analysis. The Monte Carlo simulation involved in
generating the quantities $\vec{a}^{\rm th}$ allows for possible
statistical fluctuations. Since we only simulate a finite number
of events there is always some uncertainty in the prediction.
Therefore knowing whether the differences between
Tables~\ref{tbl:sugra_chi_A} and Tables~\ref{tbl:sugra_chi_B}
and~\ref{tbl:sugra_chi_C} are significant would require generating
an ensemble of such tables. Due to the computational intensity of
such an undertaking we have not performed this analysis. When real
data exists such an extended study would be worthwhile. Here we
merely wish to examine the potential of the method.

An approximate measure of this uncertainty can be obtained by
repeating the procedure several times. For example, consider the
following mSUGRA point
\begin{equation}
\tan\beta=40 \quad m_{1/2}=380 \quad m_0=600 \quad A_0 =0 \quad
{\rm sgn}(\mu)>0.
\label{sugra2} \end{equation}
Using {\tt PYTHIA} we generated five tables such as those above
using different random number seeds in each case and $N_{\rm
event} = 50,000$ events. Using the experimental ``data'' generated
by the values in~(\ref{sugra}) we determined the value of
$\(\chi^2_T\)_m$. For this exercise we include only the LHC
collider signatures. In the limit as $N_{\rm event} \to \infty$
all five simulations should give identical answers. But the finite
value of $N_{\rm event}$ allows for some variation in the
outcomes. This variation is shown below in
Table~\ref{tbl:ensemble}.

\begin{table}[ht]
\begin{center}
\begin{tabular}{|c|c|c|c|c||c|}\hline
$\chi^2_1$ & $\chi^2_2$ & $\chi^2_3$ & $\chi^2_4$ & $\chi^2_5$ &
Mean
\\ \hline \hline 1.60& 1.48& 1.57& 1.37& 1.77& 1.56  \\ \hline

\end{tabular}
\end{center}
\caption{\label{tbl:ensemble} \footnotesize \textbf{Variation of
values for $\(\chi^2_T\)_m$ for an ensemble of simulations.}}
\end{table}

The fluctuation on $\chi^2$ is approximately within $\pm 0.2$. It
seems reasonable, therefore, to assume that the basic conclusion
from Tables~\ref{tbl:sugra_chi_A} through~\ref{tbl:sugra_chi_D} is
that while the sign of $\mu$ can be reliably determined from this
small set of experimental observables the best fit $\chi^2_T$ is
not sensitive enough to determine the value of $\tan\beta$.

\vspace{0.2in}

So far we have been working only within a given theory so we have
only been testing the rules of statistics which allow one to
extract, from a set of data, the preferred values for the
parameters of the theory. Nevertheless this exercise demonstrates
the power of the technique since measuring $\tan\beta$ or
$M_{1/2}$ by inverting formulas for cross-sections will not
succeed in the short run, while this approach might -- but the
procedure is only valuable if we already know the ``right''
theory. What would be the result if we tried the same process with
a different test theory?

We next wish to study whether the mSUGRA model itself can be
selected from among the other models of Section~\ref{sec:models}.
If so the method would become an exciting approach to how data can
guide us to theory. Our set includes several models that are
similar to minimal supergravity -- both in terms of the input
parameter set and the pattern of signatures given in
Table~\ref{tbl:sig}. In particular let us consider models F~and~J.
These are the weakly coupled heterotic model with K\"ahler
stabilization and the open string model with $D_5$ branes,
respectively.

Model~F has three free parameters: the value of the beta-function
coefficient for the condensing gauge group in the hidden sector,
the value of the gravitino mass and $\tan\beta$. Model~J is even
more constrained, having only two free parameters apart from the
sign of $\mu$: the value of $\tan\beta$ and the value of the
gravitino mass. How well can these two models reproduce the
experimental ``data'' generated from the mSUGRA
point~(\ref{sugra})? The best-fit point for the heterotic
dilaton-dominated model with K\"ahler stabilization had the
parameter values $\lbr m_{3/2} = 2750 \GeV \; , \; b_{+} =
24/16\pi^2 \; , \; \tan\beta=5 \rbr$ which corresponds to
$\(\chi^2_T\)_m = 2.8$. For the dilaton-dominated $D_5$-brane
model the best fit point was given by $\lbr m_{3/2} = 400 \GeV \;
, \; \tan\beta=10 \rbr$ which corresponds to $\(\chi^2_T\)_m =
4.5$.

In Table~\ref{tbl:compare} we collect these two best-fit points
with the four preferred points from Tables~\ref{tbl:sugra_chi_A}
through~\ref{tbl:sugra_chi_D}. Note that the various models have
different numbers of degrees of freedom. All utilize the five
collider signatures plus the non-collider measurements of $g_{\mu}
-2$ and ${\rm Br}(b \to s \gamma)$, giving a dimension of
$\vec{a}^{\rm exp}$ of seven. The number of free parameters is the
dimension of the vector $\vec{x}$, which is four for the mSUGRA
models, three for Model~F and two for Model~J. We break down the
overall contribution to the value of $\(\chi^2_T\)_m$ from each of
the types of experimental data we consider, as well as giving the
values of other kinematic data that could be used to discriminate
between models. $M_{\rm eff}^{\rm peak}$ is the peak of the
$M_{\rm eff}$ distribution where $M_{\rm eff}$ is defined as the
sum of transverse energy of all jets plus $\notE$. The quantity
$m_{ll}^{\rm peak}$ is the peak of invariant mass distribution of
the two leptons in the opposite sign dilepton events.

\begin{table}[th]
\begin{center}
\begin{tabular}{|c||c|c|c|c|c|c|} \hline
 & \multicolumn{2}{|c|}{mSUGRA $\tan\beta=10$} & \multicolumn{2}{|c|}{mSUGRA
 $\tan\beta=40$} & \multicolumn{2}{|c|}{ } \\ \hline
 &  $\quad \mu>0 \quad$ &  $\mu<0$ &  $\quad \mu>0 \quad$ &
  $\mu<0$ &  Model J &  Model F \\
  \hline \hline
LHC & 0.09 & 0.21 & 1.70 & 17.17 & 3.10& 1.7 \\ \hline
$a_{\mu}^{\SUSY}$ & 0.44 & 1.83 & 0.03 & 2.96  & 0.43& 1.0
\\ \hline
${\rm Br}(b\to s \gamma)$& 1.16 & 4.96 & 0.66 & 11.09 & 1.00& 0.1
\\ \hline
Total $\(\chi^2_T\)_m$ & 1.69 & 7.00 & 2.39 & 31.22 & 4.54& 2.8
\\ \hline\hline
d.o.f          & 3    & 3    & 3    & 3   & 4   & 5     \\ \hline
$M_{\rm eff}^{\rm peak}$ (GeV) & 1360 & 1260 & 1360   & 1438 &
1388 & 987
\\ \hline $m_{ll}^{\rm peak}$ (GeV) &92   & 92  & 92    & 92  & 92   &
58
\\ \hline
\end{tabular}
\end{center}
\caption{\label{tbl:compare} \footnotesize \textbf{Breakdown of
best-fit $\(\chi^2_T\)_m$ for different models.} Each column
represents the point in a model's parameter space which minimizes
$\chi^2_T$ in the corresponding model. The first four columns are
mSUGRA models with different choices of $\tan\beta$ and sign of
$\mu$. The fifth model is the $D_5$-brane model (Model J) and the
last one is the heterotic dilaton-dominated model (Model F). We
break down the total $\(\chi^2_T\)_m$ into contributions from the
five collider signatures (``LHC'') as well as contributions from
$a_\mu^{SUSY}$ and ${\rm Br}(b\to s \gamma)$. The kinematic
variables $M_{\rm eff}^{\rm peak}$ and $m_{ll}^{\rm peak}$ are
also given to show their utility in separating models.}
\end{table}

By comparing the $\(\chi^2_T\)_m$ with the degrees of freedom for
each model it is clear that an mSUGRA model with negative $\mu$ is
disfavored. This is mainly due to the $b \to s\gamma$ constraint,
and the discrepancy between the prediction of the theory and the
experiment result gives a large $\chi^2_T$ contribution. The
$D_5$-brane model is marginally compatible, with the primary
contribution to $\chi^2_T$ coming from fitting the LHC data. The
differences of $\(\chi^2_T\)_m$ between $\tan\beta=10$ and
$\tan\beta=40$ (both with positive $\mu$) are not significant.
This tells us that the observables we included in the $\chi^2_T$
fit are not sensitive to $\tan\beta$. Suppose at Tevatron or LHC
there is a $B_s\to\mu^+\mu^-$ signal or sensitive limit, then we
can include this observable into our $\chi^2_T$. Because of the
large sensitivity of ${\rm Br}(B_s\to\mu^+\mu^-)$ to $\tan\beta$
when the latter is large~\cite{Babu:1999hn}, such a signal can
help us discriminate between models.

The dilaton-dominated heterotic model seems to be a good fit with
the data, when one looks only at the $\chi^2_T$ variable. But
notice that the kinematic observables for this model are very
different from the other five cases. For the first four mSUGRA
models the mass difference between $\tilde N_2$ and $\tilde N_1$
is larger than the $Z$ boson mass, so the peak of the $m_{ll}$
distribution, where $m_{ll}$ is the invariant mass of opposite
sign dileptons, is around the $Z$ mass. For the dilaton-dominated
model with K\"ahler stabilization this mass difference is smaller
than $M_Z$ so the peak of $m_{ll}$ is significantly smaller than
the others. If we take this kinematic observable into account,
this dilaton-dominated SUSY breaking model should be disfavored.

Thus, in this simple example, the two mSUGRA models with positive
$\mu$ give the best fit and the other models are disfavored. This
example shows by using such a ``theory global fit'' approach, it
is possible to gain information about underlying theories.

\subsection{Identifying targeted inclusive signatures}
\label{sec:IPO}

If we focus our attention solely on the rows in
Table~\ref{tbl:sig} that we have been able to fill, we see that
some model points are indeed distinguishable. But others are not
and this differentiation is mild: many models differ from one
another by a few inclusive signatures despite large differences in
their underlying theory. If we were to subdivide the models even
more this distinguishability between representative points of the
models might be lost.

The origin of this mild differentiation can be traced to a number
of factors. Perhaps principal among them is the sheer volume of
existing data that any extension of the Standard Model must
satisfy, not to mention constraints on the masses of new particles
from direct search experiments or theoretical prejudices such as
accounting for dark matter with relic neutralinos. Theories which
{\em a priori} have quite different generic features tend to
behave quite similarly once one restricts them to a set of
parameters allowed by this data. Nevertheless, the
phenomenological approach we advocate below could serve to
estimate to what extent there is a similarity. For example, the
minimal gaugino mediation model (Model~K) has an identical set of
inclusive signatures as that of the minimal gravity mediation
model we considered (Model~A). But this similarity ultimately
stems from the fact that the GUT-nature of the gaugino mediation
demands a universal gaugino mass and scalar mass at some point
above the GUT scale (even though the actual values chosen for
these parameters in Models~A and~K are quite different). If we
were able to make real predictions for other observables we would
undoubtedly find those which would be of most use in
distinguishing between these two scenarios.

Another factor is the choice of inclusive signatures themselves.
We have chosen the entries in the collider section of the table
because these are the discovery modes that will be employed in
SUSY searches at hadron colliders and thus they have been studied
for some time. But while they may make excellent signatures for
detecting beyond the Standard Model physics, they are often not
signatures that are most directly tied to the key features of the
underlying theories themselves. For example, the supersymmetric
Higgsino mass parameter $\mu$ is a quantity whose value is
unlikely to be measured directly until a linear collider with
polarized lepton beams exists~\cite{BrKa98}. Even then extracting
the value of this parameter will be a difficult challenge as there
is no one experimental observable that is directly tied to its
magnitude. Yet the value of the $\mu$-term is perhaps the most
crucial parameter in the supersymmetric Lagrangian of the MSSM.
Knowing it's value, even approximately, might single out whole
classes of theories by pointing to a superpotential vs. K\"ahler
potential origin for this parameter and shed light on the soft
Lagrangian through the fine-tuning of the Z-boson mass.

What is needed, then, is increased study towards identifying new
signatures that more directly probe the Lagrangians of these
models without necessarily reconstructing those Lagrangian
parameters themselves. Call them ``targeted inclusive
signatures.'' They will likely not be observables that are most
efficient at discovering superpartners, but they may well prove
extremely efficient at determining which subset of high energy
theories is most likely to be correct. By constructing tables such
as Table~\ref{tbl:sig} using a variety of signatures the most
powerful discriminants (i.e. the most targeted signatures) might
become readily apparent.

It is important to keep in mind that these new signatures must be
real observables. Many analyses on how underlying theories will be
gleaned from data are based on measurements of what we will call
``in-principle observables'' (IPOs).\footnote{If this acronym
carries a certain connotation with the reader, it's intentional.}
These include soft Lagrangian parameters or ratios of parameters
that are considered key indicators of some feature of a model. But
such quantities may never be adequately measured -- at least not
to the precision that is often expected in order to distinguish
between models. Even the soft gaugino mass parameters will require
a suite of careful measurements to unambiguously
reconstruct~\cite{KnMo00}. In order to get at the correct model as
effectively as possible we must not mislead ourselves into
treating these IPOs as actual observables, at least not for a long
time. For the foreseeable future, except possibly for
(approximately) the gluino mass, the only truly measurable
quantities are the inclusive signatures. Nevertheless, the
targeted inclusive signatures with the greatest discriminatory
power are likely to be those which relate as closely as possible
to the IPOs that are so often studied in the literature.

Some reflection upon the types of models considered in
Section~\ref{sec:models} -- particularly those closely tied to
string theories -- suggests a few important distinguishing
variables. To begin with, it is apparent that the pattern of
gaugino masses is a distinctive signature of various transmission
mechanisms of supersymmetry breaking. We have every reason to
believe that more than one such mechanism (gauge mediation,
anomaly mediation, moduli mediation, etc.) is likely to be present
in nature. While one may be dominant over the others, there may be
special circumstances such as a particular point in moduli space
or a peculiar arrangement of the matter content of the theory such
that two transmission mechanisms are competitive with one another.
Furthermore, if gaugino masses are small at the tree level for one
reason or another, then it is likely that the leading order
gaugino masses will be nonuniversal. Thus a useful IPO would be
the ratio
\begin{equation}
\order_1 \equiv \frac{M_2 - M_1}{M_3} .
\label{IPO1} \end{equation}

Another parameter of critical importance from a theoretical
standpoint is the value of the fundamental mass scale of the
theory. Here by fundamental scale we mean the scale at which the
input parameters of the MSSM are set. In a string context we might
think of this scale as the string scale. Though it does not have
to be the case, it is typical to assume that this fundamental
scale is to be identified with the scale of supersymmetry breaking
in the observable sector. It is then often further assumed that
this high-energy input scale is the GUT scale. Some indication of
the magnitude of this fundamental scale can be obtained from the
pattern of scalar masses at the low scale, in particular the
difference between typical masses in the squark sector versus the
slepton sector. Given that squark masses are generally strongly
influenced over the course of their RG evolution by the gluino
mass while slepton masses run very little, some measure of the
size of the RG ``desert'' can be related to an IPO such as
\begin{equation}
\order_2 \equiv \frac{m_{\tilde{t_1}}^{2} -
m_{\tilde{\tau_1}}^{2}}{M_3^2} .
\label{IPO2} \end{equation}
Of course tree-level non-universalities in scalar masses at the
input scale would obscure the impact of the RG running, but many
models that predict such non-universality (such as gauge and
anomaly mediated models) often involve characteristic splittings
between squarks and sleptons that would make a quantity such
as~(\ref{IPO2}) a useful parameter.

Even the simple ratio of gaugino masses to scalar masses alone
will be of critical importance in understanding the nature of
supersymmetry breaking in a hidden sector and transmission to the
observable sector. In minimal supergravity both scalar and gaugino
masses are treated as free input parameters, but more
sophisticated models often make predictions about the relative
sizes of these masses since one may be generated at tree level
while the other arises only at the loop level. Within the scalar
sector itself the presence of large mass splittings in a sector
other than the stop sector might reflect important information
about scalar mass non-universality as well as the possible field
dependence of trilinear coupling which appear in the off-diagonal
entries of the scalar mass matrices. A potentially useful IPO
might be the ratio
\begin{equation}
\order_3 \equiv \frac{m_{\tilde{{\tau}_2}}^{2} -
m_{\tilde{\tau_1}}^{2}}{M_1^2} .
\label{IPO3} \end{equation}

As important as these IPOs are for understanding the structure of
the underlying supersymmetric theory, they are not in themselves
directly measurable. Translating these quantities into inclusive
signatures will necessarily obscure their relation to the
fundamental Lagrangian somewhat. A good targeted inclusive
signature would be one that is shown, through event simulation in
real collider environments, to correlate well with the underlying
IPO. Our efforts to identify such signatures based on the collider
observables of Table~\ref{tbl:sig} have thus far produced only
limited success. The primary obstacle is the inclusive nature of
the signature: when an observable arises through a combination of
channels a strong correlation between that observable and the IPO
may exist only for a particular hierarchy of masses or over a
particular range of parameters.

Unfortunately, actually measuring any such observable may be very
difficult. We have not found any such observables that are likely
to be measurable before a linear collider exits, whenever that may
be. Nevertheless, we are hopeful that further phenomenological
study will produce targeted inclusive signatures with robust
correlations to key soft Lagrangian combinations such
as~(\ref{IPO1})~-~(\ref{IPO3}) and encourage others to suggest
such observables. One avenue that may prove fruitful is the use of
layered, or sequential, targeted signatures. That is, using the
measured value of one inclusive signature as a key to indicate
what subset of targeted observables will correlate well with the
data.

We conclude this section by noting that the global inclusive fit
to theory technique of the previous section may also lead in an
indirect way to the targeted signatures we seek. The theories we
study are defined by an input parameter vector $\vec{x}$ as in the
case of~(\ref{xsugra}) above. This defines a vector space that is
the appropriate one for studying the underlying theory. But the
inclusive signatures themselves depend on the low-scale values of
the soft parameters. The mapping from the high to low scale is
performed using the RGEs. Let us call the low-scale vector of soft
parameters $\vec{y} = \vec{y}(\vec{x})$. This mapping is not
one-to-one, but the elements of $\vec{y}$ are related in a known
way.

\begin{figure}[thb]
\begin{center}
\includegraphics[scale=0.6,angle=270]{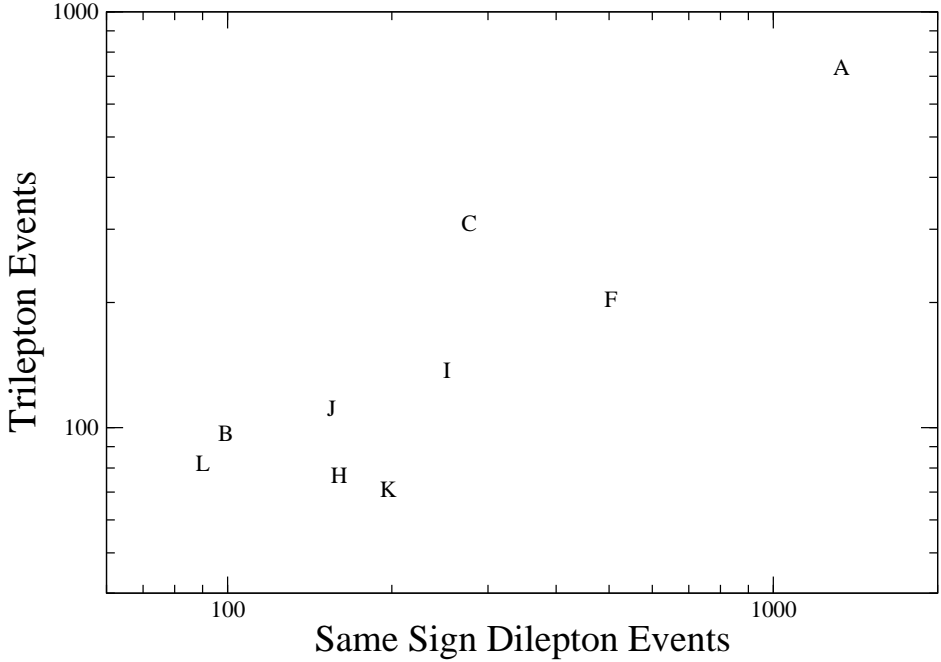}
\caption{\footnotesize \textbf{Distribution of Models.} This
figure illustrates how different models of high-energy physics
tend naturally to produce different types of inclusive signatures.
Combining such figures for several signatures could provide
significant discriminatory power.} \label{fig:ratio1}
\end{center}
\end{figure}

The theory $\chi^2$ variable $\chi^2_T$ is a function of the
original high-scale soft parameters only indirectly through the
object $\vec{y}$. It is also a function of the observables we
choose to include in the ``experimental'' result $\vec{a}^{\rm
exp}$
\begin{equation}
\chi^2_{T} = f(\vec{y}(\vec{x}); \, \vec{a}^{\rm exp}) .
\label{fchi} \end{equation}
The targeted IPOs of Section~\ref{sec:IPO}, such as those
suggested in equations~\ref{IPO1},~\ref{IPO2} and~\ref{IPO3},
represent a (non-linear) change of variables from the original
vector $\vec{y}$ to a new vector $\vec{y}\,'$. Similarly, targeted
inclusive signatures would represent new observables $\vec{a}^{\rm
exp}\,'$ built from combinations of those in Table~\ref{tbl:sig}
(such as the ratios in Figure~\ref{fig:ratio1}) -- or as yet
un-thought-of observables that can be measured. Finding
combinations of $\vec{y}\,'$ and $\vec{a}^{\rm exp}\,'$ such that
the functional form represented in~(\ref{fchi}) is as strong as
possible means looking for combinations where the gradients along
any particular component in $\vec{y}\,'$ are as large as possible.
This is essentially an algorithm for finding the optimal targeted
observables and targeted IPOs, and may prove an invaluable tool in
guiding us to measurements that will unravel the inter-related
parameters of the supersymmetric Standard Model.

\subsection{Comments for Experiments}
\label{sec:exp}

As signals for physics beyond the Standard Model begin to emerge,
particularly in collider data at the Tevatron or LHC, our approach
has some impact on how data should be treated. Of course it will
be important to learn which superpartners are being produced, to
measure their masses, production cross sections and branching
ratios, and to deduce Lagrangian parameters if possible. To do
that, normally selections and cuts are performed on data to reduce
backgrounds and isolate signals. At the Tevatron, with limited
statistics, that procedure may reduce the signal so much that
little can be learned. At LHC the large number of channels may
make separation of states very difficult. Even if that were
possible, at hadron colliders there are in general fewer
observables than relevant Lagrangian parameters, so learning the
essential Lagrangian quantities may not be possible except in
special lucky situations.

Our approach has implications for these issues. We argue that the
mere existence of certain classes of events -- and the relative
amounts of these different classes -- point toward some classes of
models and not others in useful ways. Thus experimenters should
attempt fully or nearly inclusive measurements, without cuts that
reduce statistics. It is, of course, essential to know the
Standard Model predictions very well in order to recognize when
they are exceeded.

\section{Comments for Model Building}
\label{sec:mod}

It is both ironic and disappointing that the majority of the
``predictions'' being made by the bulk of SUSY models in the
literature are in the one sector where we currently have precisely
zero data: superpartner masses. Meanwhile we are blessed with
large amounts of data from the other sectors displayed in
Tables~\ref{tbl:sig}. The measurement of the Z-boson mass alone
tells us a great deal about EWSB and how the $\mu$-term might be
generated~\cite{KaLyNeWa02}. Electroweak precision measurements
favor values of the oblique parameters that nearly (but perhaps
not precisely) coincide with the Standard Model values, which
tells us a great deal about any acceptable theory. Measurements at
the Tevatron and the B-factories are providing important data on
rare decays, mixing matrix entries and CP violation. The recent
measurement of neutrino oscillations provides a whole sector
awaiting a supersymmetric explanation. Cosmological observations
have become sufficiently precise to rival terrestrial limits in
some areas and are giving evidence for new classes of particles
such as cold dark matter relics and quintessence fields. Indeed, a
cosmological Standard Model is taking shape that cries out for
(supersymmetric) explanations for dark matter, inflation and
baryogenesis. Even null results in searches for certain
theoretically well-motivated new fields as axions,
fractionally-charged particles and $Z'$-bosons provide important
constraints on new models.

To be fair, there are a great many supersymmetry-based models in
the literature that address each of these issues. But all too
often this is done in isolation, with little or no regard to how
the issues raised in say, rare decays may restrict the solutions
at our disposal for understanding the origin of Yukawa textures or
SUSY breaking in a hidden sector. It is our contention that the
model builder must seek to treat the entire arena of inclusive
signatures in as comprehensive a way as possible. We cannot
completely neglect the theoretical framework in which these models
arise -- they cannot truly be reduced to the MSSM with an
mSUGRA-like soft Lagrangian. The theory elements used to solve the
various SUSY ``problems'' will manifest themselves in different
ways depending on the construction. That is, there will always be
some ``back reaction'' between, say, a $\mu$-term generating
mechanism, or a flavor texture, on the low-energy predictions of
the model. Thus dealing successfully with some problem is only an
initial step towards solving the problem.

For example, models that utilize gauge mediation to communicate
supersymmetry breaking to the observable sector have a
particularly severe $\mu$ problem, or more specifically a $B\mu$
problem: it is difficult to engineer a $\mu$ and a $B\mu$ term
that are both of electroweak scale size~\cite{GiRa99}. But like
all such problems, many solutions have been proposed. If the $\mu$
term is generated by a higher-derivative operator as in the
mechanism of~\cite{DvGiPo96} the problem can be solved, at the
expense of adding two new singlets to the theory. This mechanism
is economical and holistic in that the $\mu$ and $B\mu$ terms
arise from the same mechanism as the other soft parameters in the
theory. However, the Higgs soft scalar masses are also modified
from the predictions of the minimal gauge mediated theory. This
will, of course, affect the low energy phenomenology in a
model-dependent way.

Another example involves combining a mechanism of generating
Yukawa textures and supersymmetry breaking within an integrated
string-based model~\cite{DuGrPoSa96}. If we treat Yukawa textures
in isolation then there are a great many that have been designed
to reproduce the correct fermion masses and mixings. In
particular, it is possible to generate the appropriate textures
using only a single Froggatt-Nielsen field $\varphi$ charged under
a single Abelian $U(1)$ flavor group with a vacuum value $\lang
\varphi \rang \simeq \theta_{c} \simeq
0.2$~\cite{LeNiSe93,LeNiSe94,LeTr98}. This single $U(1)$ mechanism
works well when soft supersymmetry breaking can be ignored, as in
the universal scalar mass paradigm. But since this $U(1)$ is
necessarily broken by the vacuum value of $\varphi$, it is
possible -- even likely -- that it will generate D-term
contributions to the soft masses of the various Standard Model
fields that are flavor dependent. Clearly, this changes the
phenomenology of the model. In particular, it now becomes
difficult to suppress the mixing in the $K-\bar{K}$ sector without
resorting to two Abelian flavor factors. The flavor sector has
affected the soft term phenomenology, which in turn has required a
change in the flavor sector. Even if these D-term contributions
could be eliminated, the model independent supergravity
contribution to the trilinear A-terms proportional to $\lang
\varphi \rang m_{3/2}$ will provide non-universal contributions to
the A-terms that are not proportional to the original Yukawa
couplings~\cite{DuGrPoSa96}. The question of whether these
contributions are phenomenologically dangerous now depends on the
exact form of the superpotential, the soft Lagrangian and the size
of the gravitino mass.

This is more than an academic exercise: most models that receive
phenomenological study are those that are promoted on the basis of
their perceived predictivity. That is to say, they are promoted on
the basis of their apparent simplicity and the small number of
free input parameters that they seem to have. But this is
deceptive, as the examples above suggest. The question of how well
these simple ``paradigm models'' approximate the realistic models
we actually need to construct is an open question.

What is more, the way that experimental measurements are used to
constrain models depends very much on the properties of a complete
model. For example, if we choose to neglect the flavor sector of a
model by tacitly assuming that degeneracy or alignment is at play,
then we may use the minimal flavor violation model to calculate
the implication of the CLEO and BELLE measurements of ${\rm Br}(B
\to s \gamma)$ on the particle masses and the sign of the $\mu$
parameter of a given model. But when there are new sources of
flavor violation in the soft Lagrangian, many other contributions
must be included in the calculation which are not suppressed by
CKM matrix elements~\cite{Gabbiani:1996hi}. It is even possible
for this new contribution to be the dominant one, thus changing
the interpretation of the experimental result~\cite{EvKaRiWaWa02}.
A completely different set of constraints on the superpartner
spectrum, and no effective constraint on the sign of $\mu$, is
obtained. So without a complete model this important clue to the
supersymmetric world cannot be properly utilized. A similar thing
occurs in the neutralino cold dark matter scenario, where the
interaction cross section of relic neutralinos with protons (as
well as the relic density itself), depends crucially on the phases
in the neutralino mass matrix~\cite{BrChKa01}. How are we to
properly interpret the implication of a dark matter detection
signal -- or apply the WMAP constraint on non-baryonic dark matter
-- without a theory that addresses the origin of $\mu$ and its
relative phase with respect to gaugino masses?

The answer is, of course, quite familiar. We make assumptions that
are simplifying in order to separate confounding issues and make
progress. If these assumptions were justified then they might be
put to the test at a later time with a different set of
experiments -- if the assumptions are very wrong this analysis is
irrelevant. But without examples of complete models it is unclear
what these cross-checking experiments might be. We suggest that in
constructing models that are capable of filling in all the entries
in tables such as Table~\ref{tbl:sig} these relations will become
much more obvious.

What might a complete model look like? The promise of
supersymmetry is the opportunity to make meaningful statements
about mass scales in the effective field theory below the string
scale by protecting these scales from quadratic divergences. Thus
a supersymmetric Standard Model should explain the dynamical
origin of the mass scales relevant to observations: the
supersymmetry breaking scale in the hidden sector, the scale of
transmission of that breaking to the observable sector, the value
of the $\mu$ parameter, the size of the Higgs vev, the axion decay
constant, Majorana neutrino masses, the scale at which flavor
symmetries are broken, etc. While conceptually distinct from
issues of supersymmetry breaking and mass scales, a supersymmetric
Standard model -- to the extent that it is a deeper model than the
non-supersymmetric Standard Model -- should also identify the
symmetries (if any) operative in the dimensionless parameters of
the effective theory below the string scale. The origin of these
parameters and the symmetries themselves are presumed to be in the
domain of a future stringy Standard Model. Finally, the SSM we
seek will be consistent with the constraints of
Table~\ref{tbl:sig} but will also provide an explanation for why
this consistency {\em is an inherent feature of the model itself}
by deriving this fact from deeper physics principles.

For example, we might ask what it might require to make mSUGRA a
complete model. In truth it is already complete in some very
limited sense: namely that the assumptions made in defining the
mSUGRA paradigm are technically sufficient to fill in all of the
entries of Table~\ref{tbl:sig}, though some in a very trivial way.
Let us call this ``minimal completeness:'' mSUGRA is complete, but
not truly a {\em model} because it does not contain an explanation
for its various assumptions. In mSUGRA this comes about by
postulating some hidden sector superfields which develop vacuum
values for their auxiliary field components in a universal way.
Yukawa matrices and the $\mu$ parameter are taken as
(non-field-dependent) fixed parameters. The neutrino sector and
cosmological issues are ignored. While this set of assumptions is
technically sufficient to fill most of the entries in
Table~\ref{tbl:sig}, it clearly provides no explanation of deeper
physics principles and can only be thought of as a framework
deserving of its title as ``minimal.''

To make mSUGRA complete in the sense we use the term would require
adding several elements to this minimal framework. The origin of
these auxiliary field vacuum values would need to be explained and
the vanishing vacuum energy outcome demonstrated through the
construction of a concrete dynamical symmetry breaking mechanism.
The reality of the soft terms should be traced to some symmetry or
some mechanism for predicting the imaginary parts of the relevant
vacuum expectation values. The $\mu$-term would be generated
dynamically, either through some sort of Giudice-Masiero mechanism
or superpotential vev so that its value could no longer be taken
as an input but instead be tied to other parameters in the theory.
The smallness of $M_Z$ relative to the soft Lagrangian parameters
would then be explained in terms of this mechanism. The neutrino
sector would need to be added. If the seesaw mechanism is to be
employed then the scale of the Majorana masses should be derived
from the dynamics of some field. If flavor symmetries are at play
in the Yukawa matrices, broken by the vacuum values of some set of
flavon fields, then these vevs must be understood in terms of
internal dynamics of the model -- even if the symmetries or flavor
charges themselves are taken as inputs. Since the region of
parameter space where electroweak baryogenesis can occur while
simultaneously satisfying LEP constraints on superpartner masses
requires non-universal soft terms, the mSUGRA model would need to
be augmented with some CP violation and additional fields and
couplings beyond the MSSM. This list could continue, but it is
already clear that while the simple structure of soft terms that
characterizes the mSUGRA paradigm could be retained, it would need
to be embedded in a much larger theoretical structure that would
involve many new internal relations that might severely restrict
the allowed parameter space we typically associate with this
model.

\section{Suggestions for String Theorists}
\label{sec:string}

And how can formal theorists and string phenomenologists help in
our endeavor to discover the supersymmetric Standard Model? As we
begin the arduous process of building realistic models from the
wealth of ideas that have been studied so far we will find our
models will never be truly complete. We may strive to find
explanations for the phenomena we see around us, and couch that
explanation in dynamics of a supersymmetric field theory, but many
of the initial conditions that seem to generate the desired
outcomes will remain inputs to the theory. Thus the particle
content, gauge groups, charge assignments and so on that explain
flavor structures or neutrino masses or gauge coupling unification
becomes the output of a future string supersymmetric Standard
Model (SSSM). Relations among soft terms or symmetries in Yukawa
couplings may be reflective of specific moduli sectors or may
arise only in specific types of compactifications. As model
builders begin to understand what pieces work well together it
should become possible to loosely associate classes of
supersymmetric models with classes of string constructions.

In the meantime there are ways that model-builders, especially
those who focus on string-based models, can be assisted in making
more efficient choices in the models they study. String theorists
could take more interest in the phenomenology done in their name;
not only by listening to and discussing attempted analyses
constructively, but also by being ready to step in when the models
studied are not as reflective of the state of current string
knowledge as they could be. Within any class of string theories,
from weakly coupled heterotic strings to open strings at strong
coupling, the key determinants of the low energy phenomenology are
the moduli dependence of the K\"ahler potential, superpotential
and gauge kinetic functions. Knowledge of these functions at the
tree level and loop level, where possible, is the starting point
all of subsequent analysis. An increasingly general classification
of models on the basis of moduli taxonomy would serve as a useful
guide to low energy model building. More specifically, questions
can be formulated whose answers are important for string
phenomenology and whose answers have to come from string
theorists~\cite{questions}.

In addition, crucial aspects of low energy phenomenology often
depend on the values of numbers and charges that have a
string-theoretic origin: modular weights, anomalous U(1) charges,
oscillator numbers, anomaly cancellation coefficients and
topological quantities to name a few. Knowing even general ``rules
of thumb'' about how typical values of these numbers are related
to particular constructions or can be associated to particular
fields in the observable sector would be of great value in
choosing the most promising avenue for top-down study.

The notion of associating classes of string constructions to
classes of four-dimensional supergravity theories with distinct
properties complements nicely the association of theories to
signatures we seek to promote in this paper. Indeed some work has
already begun in this direction~\cite{Do03}. In tandem, these
approaches may allow a general connection between forthcoming data
and string models to be formed.

\section{Conclusion: How will a Supersymmetric Standard Model Emerge?}

Let us clearly state from the outset that we do not claim to have
answers about what the supersymmetric Standard Model will look
like. Nor have we yet incorporated all the suggestions we make
here into our own thinking. But we feel that the time is right to
begin speculating on how a supersymmetric Standard Model might
emerge and what strategies can be taken now to prepare for the
supersymmetric era.

To date most analyses that connect high scale theories to data
have been of the top-down variety. All benchmark studies are of
this variety and the analysis incorporated in Table~\ref{tbl:sig}
is an example. This is a necessary exercise that is invaluable in
increasing our intuition on how theories reflect themselves in
data. But as we noted above, contact is currently being made only
in isolated patches of the complete set of inclusive signatures
that we have at our disposal today. This is in large part due to
the incomplete nature of the models that are available in the
literature and to a lesser degree the result of specialization
among high energy theorists that tends to treat the various groups
of signatures as separate domains with their own analysis tools,
effective Lagrangians and simplifying assumptions. Expanding these
models to a greater degree of realism by combining ideas from many
sectors -- fermion masses, neutrino mixings, the $\mu$ term and
axions, baryogenesis, extended gauge groups, etc. -- will likely
require changing the basic features of the model itself and thus
the superpartner spectra and collider signatures as well. This may
be seen as an obstacle to making robust predictions or as an
opportunity to better ascertain which frameworks are flexible
enough to have hope of encompassing the eventual supersymmetric
Standard Model.

Reconstructing the ``right'' model armed only with benchmark
studies will obviously be difficult. If nature is kind we may find
the set of inclusive signatures (observations vs. non-observations
of new physics in various channels) that are uncovered by
experiment match quite well with one of the familiar models of
today. Presumably we can only increase the odds of this occurring
by expanding the number of models under consideration and aiming
for a continuum of signatures. But this quite rapidly reaches a
point of diminishing returns and is hardly the most efficient
manner in which to seek the supersymmetric Standard Model.

If we are patient we might consider accumulating observations over
time. As signatures are observed they become constraints, and for
any particular model a locus of points consistent with these
constraints can be identified. In principle this method can
eventually eliminate the entire parameter space of a model --
allowing us to discard it and move on to the next. But in truth
this requires a great deal of time and iteration: as parameter
space is carved away model builders will make small changes to the
base framework, enlarging the parameter space only to have
additional observations constrain it again. Such a process is
already beginning with the minimal supergravity paradigm as the
constraints of ${\rm Br}(b \to s \gamma)$ in the minimal flavor
violating paradigm, the WMAP measurement of the cold dark matter
abundance and the current LEP limit on the lightest Higgs boson
mass are used to pare down the allowed parameter space. But this
only motivates small deviations from mSUGRA, such as nonuniversal
scalar masses or non-universal gaugino mass relations. If we
already had reason to believe that the model in question (mSUGRA
in the example above) was likely to be a strong candidate for a
supersymmetric Standard Model then this would not necessarily be
an inefficient way to proceed. Indeed, once a putative SSM is
agreed upon the process of measuring the complete soft Lagrangian
and overconstraining the parameter space of that model will begin
-- ultimately bolstering that model's claim as the SSM or
disproving it in favor of a variant. This is the status of the
non-supersymmetric Standard Model today.

This process could be rapidly accelerated in two different ways.
First, if the models we consider were truly complete then all of
the data at our disposal could be brought to bear simultaneously.
When forced to this degree of realism it is likely that few
candidate models could survive even the first hints of data at the
beginning of the data era. In fact, it is unlikely that many of
the currently existing models would satisfy even the constraints
that we {\em already possess} on any putative SSM. Are the
inclusive signatures included in Table~\ref{tbl:sig} infallible?
Of course not -- each of these models has a parameter space
associated with it that is consistent with the constraints of
Table~\ref{tbl:sig} and, depending on the model, the derived
constraints as well. In some corners of that parameter space one
or two of the expected observation modes may disappear, or new
observation signatures arise. But this only underscores the need
to study a broad spectrum of theories and include an ever-wider
array of inclusive signatures in our analysis.

Second, clever thinking about new sorts of inclusive signatures
(or combinations of inclusive signatures from different types of
experiments) may provide more targeted information that can go
directly to the key differentiating feature of high energy models.
It is imperative that we keep focused on the broad nature of
inclusive observables as opposed to the eventual measurement of
``in-principle observables'' (IPOs). The most useful observables
will be those that are shown, through Monte Carlo simulation, to
correlate well with some property of the soft supersymmetry
breaking Lagrangian over a broad class of theories and wide range
of parameters within each theory. Some tentative suggestions in
this directions were given here. We have no doubt that much better
ideas can be found in the near future.

The theory global fit method we propose can be a useful
stepping-stone from the first experimental evidence of the
supersymmetric world to an eventual supersymmetric standard model.
In this approach one requires simultaneous consistency with
several pieces of information -- collider data as well as
non-collider observations. Each class of models will have a
parameter space over which a measure of the goodness-of-fit
$\chi_T^2$ can be constructed. Such a theory global fit will
select the most preferred set of parameters of a given class, but
it may also provide discrimination between classes when the values
of $\chi_T^2$ are compared for different theories.

The utility of this method will certainly be at its greatest in
the beginning of the data-rich era: as we are able to extract the
soft Lagrangian parameters themselves the technique will have
diminished value in finding the broad features of a model. But it
then becomes a tool for testing the robustness of a candidate SSM.
We can perturb the model in controlled ways (perhaps adding a
sector in the desert) and see how the fit changes. In this the
global fit technique is similar to studying fine-tuning (another
well-defined quantitative measure whose value is greatest when
data is most lacking) in that there are always possible caveats,
but perhaps most of them can be examined in a controlled way.

Ultimately, however, history implies that carving up parameter
space and top-down analysis is not how grand new paradigms are
unearthed. At some point in the coming data era the supersymmetric
Standard Model will be guessed through an act of intuition. This
candidate SSM will make predictions for yet unmeasured inclusive
signatures and yet unseen new phenomena. The discovery of such
phenomena, or lack thereof, will bolster the acceptance of the
model or cause it to be abandoned for a rival model. Our intuition
can only be improved if we begin to treat theories in a holistic
manner, strive to predict ever larger sets of inclusive signatures
and find ways to measure quantities that can provide the maximum
discrimination between models without complete measurements of the
soft Lagrangian parameters themselves.


\section*{Appendix}

In this Appendix we explain the inclusive signatures in some
detail and explain the meaning of ``Y'', ``$\SM$'' or $\checkmark$
in each case.

The inclusive signatures are divided into five categories, with
the first group being the LHC collider signatures. ``Large
$\notE$'' means $\notE \ge 100 \GeV$. A prompt photon signature is
defined theoretically as a model for which the NLSP is the
second-lightest neutralino $\wtd{N}_2$ and decays dominantly via
the mode $\wtd{N}_2 \to \wtd{N}_1 \gamma$ within the detector, or
a model for which the NLSP is the lightest neutralino $\wtd{N}_1$
which decays via $\wtd{N}_1 \to \wtd{G} \gamma$ in the detector,
as in some gauge-mediated models. Experimentally it is defined by
a high energy isolated photon with an energy above 20 GeV. The
isolated pion signature represents the case for which the lightest
chargino $\wtd{C}_1$ decays via the process $\wtd{C}_1 \to
\wtd{N}_1 \pi^{\pm}$, as in models with a high degree of
anomaly-mediation. For this signature to provide a reasonable
signal we require the transverse momentum of the pion to be
greater than 2 GeV. The trilepton, same-sign and opposite-sign
dilepton signatures are defined as in~\cite{Baer:1995va}. We
designate a model as providing a ``$\tau$-rich'' or ``$b$-rich''
signature whenever the superpartners $\tilde{\tau}$ and/or
$\tilde{b}$ are sufficiently light that gauginos will dominantly
decay into cascades involving these particles. An excess number of
jet events where the jets are tagged as having originated from a
$\tau$ or a bottom quark would constitute a beyond the SM
signature. Finally, a ``long-lived (N)LSP'' is one whose lifetime
is such that it traverses the detector before decaying.

The second category of inclusive signatures is related to issues
of flavor physics and CP violation. The first observable in this
category is $g_\mu-2$. Models with a $2\sigma$ deviation from the
SM prediction (with a positive sign) are denoted by a ``Y.'' For
the total branching ratios of $B_s \to \mu^{+} \mu^{-}$ and $B \to
X_s \gamma$, a ``Y'' in the first case represents a total rate for
this process in excess of the SM prediction. For the second case
the ``$\checkmark$'' represents a choice of parameters such that
the total SM + SUSY contribution to the rate is in accordance with
experimental observation to within $3\sigma$. For the other
inclusive signatures ``$\SM$'' implies no contribution to the
observable beyond that of the Standard Model. For the case of
neutrinoless double beta-decay no models make any predictions on
the nature of neutrino masses. If they did, however, the row would
contain a ``Y/N'' entry for each model indicating whether the
current generation of neutrinoless double beta-decay searches
would be sensitive to a signal.

The third category is related to cosmological observables.
Assuming that the LSP is the cold dark matter of our galactic
halo, then if a signal is expected by the next generation direct
WIMP detectors, the corresponding models are denoted by a ``Y.''
If space-based experiments, such as the HEAT experiment, are
expected to measure an excess above the SM prediction in the
composition of cosmic rays the model is given a ``Y'' in this
category~\cite{KaWaWa02}. Models that can give such a signal are
denoted by a ``Y''. The LSP of some models can be easily captured
in the sun and the resulting annihilation among LSPs can generate
a neutrino flux. If such a flux is detectable, the corresponding
entry should be ``Y.'' We note that we always normalize the LSP
density to the halo CDM density. Thus if the thermal relic density
of LSPs is small enough that thermal production mechanisms in the
early universe are insufficient to generate this halo density, we
then assume there is sufficient non-thermal production to make up
the required halo dark matter density. These cases are indicated
with an asterisk. None of these models includes an axion sector,
so there are no ``Y/N'' predictions for the current round of
cryogenic axion detectors. If these models were capable of making
a prediction for the baryon asymmetry of the universe they would
receive a $\checkmark$ if they were engineered to produce the
observed asymmetry. Since no such framework is provided, this row
is left blank.

In the EWSB sector, a $\checkmark$ under $M_Z$ would indicate a
model for which the Z-boson mass was correctly predicted in terms
of the values of $\mu$, $B$ and the Higgs boson soft scalar
masses. Since $M_Z$ is, in fact, an input in all these models,
this row is blank. However all models successfully fall into the
region allowed by electroweak precision measurements and receive a
$\checkmark$ in this row. As for the unification section of the
table, all but one of the models take $\alpha_s(M_Z)$ as an input
parameter, so the reproduction of the measured value of this
parameter becomes a trivial accomplishment and the row is left
blank. It should be noted that the model
of~\cite{BlDeRa02a,BlDeRa02b} performs a fit to $\alpha_s (M_Z)$
so that a prediction of some sort is made. Nevertheless the
difference between the high scale inferred $\alpha_s$ and the
value of $\alpha_1 = \alpha_2$ is then interpreted as a threshold
correction of unknown origin. All models are consistent with the
extrapolated unification of gauge couplings at some high scale. No
model makes a meaningful prediction about the rate of proton decay
so this row is also left blank. However, at least one model
postulates the existence of additional particles beyond those of
the MSSM at the TeV scale, yielding a ``Y'' in this row. In this
case these are additional leptons designed to allow for gauge
coupling unification.


\end{document}

%% file: basic.tex
\DeclareMathAlphabet   {\mathsc}{OT1}{cmr}{m}{sc}

\def\[{\left [}
\def\]{\right ]}
\def\({\left (}
\def\){\right )}
\newcommand{\lang}{\left\langle}
\newcommand{\rang}{\right\rangle}
\newcommand{\lbr}{\left\{}
\newcommand{\rbr}{\right\}}
\newcommand{\oline}[1]{\overline{#1}}

\newcommand{\wtd}[1]{\widetilde{#1}}

\newcommand{\notE}{E\kern-0.6em\hbox{/}\kern0.05em_T}
\newcommand{\notCP}{CP\kern-0.6em\hbox{/}}

\newcommand{\GeV}      {~\mathrm{GeV}}
\newcommand{\TeV}      {~\mathrm{TeV}}

\newcommand{\SM}       {\mathsc{sm}}
\newcommand{\CP}       {\mathsc{cp}}

\newcommand{\UV}       {\mathsc{uv}}
\newcommand{\GS}       {\mathsc{gs}}

\newcommand{\GUT}      {\mathsc{gut}}
\newcommand{\STR}      {\mathsc{str}}
\newcommand{\SUSY}     {\mathsc{susy}}

\newcommand{\order}{{\cal O}}

\newcommand{\gappeq}{\mathrel{\rlap {\raise.5ex\hbox{$>$}}
{\lower.5ex\hbox{$\sim$}}}}
\newcommand{\lappeq}{\mathrel{\rlap{\raise.5ex\hbox{$<$}}
{\lower.5ex\hbox{$\sim$}}}}